\documentclass{vldb}
\usepackage{graphicx}
\usepackage{balance}  
\usepackage[ruled,linesnumbered,vlined]{algorithm2e}
\usepackage{listings}
\usepackage{color}
\usepackage{multirow}
\usepackage[caption=false, font=footnotesize]{subfig}
\usepackage[htt]{hyphenat}

\definecolor{dkgreen}{rgb}{0,0.6,0}
\definecolor{gray}{rgb}{0.5,0.5,0.5}
\definecolor{mauve}{rgb}{0.58,0,0.82}

\newtheorem{theorem}{Theorem}
\newtheorem{corollary}{Corollary}
\newdef{definition}{Definition}

\vldbTitle{Kaleido: An Efficient Out-of-core Graph Mining System on A Single Machine}
\vldbAuthors{Cheng Zhao, Zhibin Zhang, Peng Xu, Tianqi Zheng, Xueqi Cheng}
\vldbDOI{https://doi.org/TBD}

\begin{document}


\title{Kaleido: An Efficient Out-of-core Graph Mining System on A Single Machine}



%
%
%
%

\numberofauthors{5} 

\author{
%
%
\alignauthor
Cheng Zhao\\
       \affaddr{ICT, CAS}\\
       \email{zhaocheng@ict.ac.cn}
\alignauthor
Zhibin Zhang\\
       \affaddr{ICT, CAS}\\
       \email{zhangzhibin@ict.ac.cn}
\alignauthor 
Peng Xu\\
       \affaddr{ICT, CAS}\\
       \email{xupeng@ict.ac.cn}
\and  
\alignauthor
Tianqi Zheng\\
       \affaddr{ICT, CAS}\\
       \email{zhengtianqi@ict.ac.cn}
\alignauthor
Xueqi Cheng\\
       \affaddr{ICT, CAS}\\
       \email{cxq@ict.ac.cn}
}

\maketitle

\begin{abstract}
Graph mining is one of the most important categories of graph algorithms.
       However, exploring the subgraphs of an input graph produces a huge amount of intermediate data.
       The ``think like a vertex'' programming paradigm, pioneered by Pregel, cannot readily formulate mining problems, which is designed to produce graph computation problems like PageRank.
       Existing mining systems like Arabesque and RStream need large amounts of computing and memory resources.

       In this paper, we present Kaleido, an efficient single machine, out-of-core graph mining system which treats disks as an extension of memory.
       Kaleido treats intermediate data in graph mining tasks as a tensor and adopts a succinct data structure for the intermediate data.
       Kaleido utilizes the eigenvalue of the adjacency matrix of a subgraph to efficiently solve the subgraph isomorphism problems with an acceptable constraint that the vertex number of a subgraph is less than 9.
       Kaleido implements half-memory-half-disk storage for storing large intermediate data, which treats the disk as an extension of the memory.
       Comparing with two state-of-the-art mining systems, Arabesque and RStream, Kaleido outperforms them by a GeoMean \textbf{12.3$\times$} and \textbf{40.0$\times$} respectively.

\end{abstract}

\section{Introduction}\label{sec:intro}

Graphs data is ubiquitous in a broad range of fields such as social networks, web networks, financial networks, biological networks, and the analysis of graphs is becoming increasingly important. Generally, we divide graph analysis problems into two major types, graph computation and graph mining. Graph computation aims to compute some meaningful values of vertices in a graph. For example, we calculate the PageRank \cite{Page:PR} value of a web graph to obtain the top-k valuable web pages; give two vertices in an input graph, we calculate the shortest path between them. While graph mining aims to discover structural patterns to meet the user's interest criteria. For example, we mine frequent subgraphs in the biological data to discover highest gene expression \cite{Hu:MiningCD}; we extract the frequency distribution of all motifs that occur in PPI network \cite{prvzulj:biological}; we discover cliques in financial networks to detect frauds\cite{eberle:insider}.

\subsection{Problem statement}

Graph computation problems can be represented through linear algebra over an adjacency matrix based representation of the graph.
Many practical solutions, like PowerGraph \cite{gonzalez:powergraph}, Ligra \cite{shun:ligra}, GraphX \cite{gonzalez:graphx}, Chaos \cite{roy:chaos}, Gemini \cite{zhu:gemini}, etc., follow a simple ``think like a vertex (TLV)'' programming paradigm pioneered by Pregel \cite{malewicz:pregel}, which is a perfect match for linear algebra problems.

\begin{figure}
       \centering
       \includegraphics[width=0.5\textwidth]{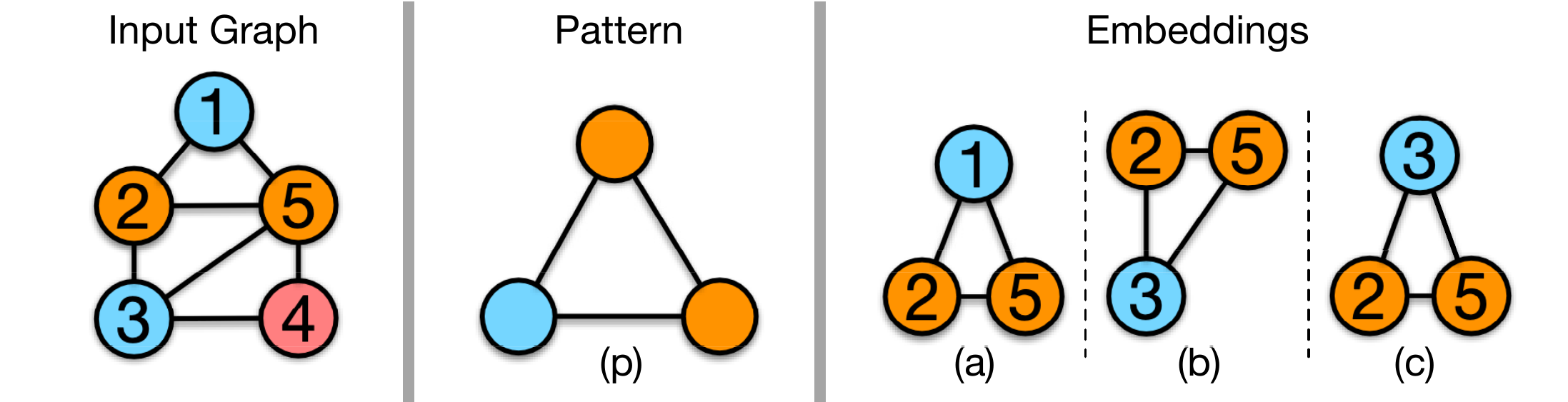}

       \caption{An example of pattern matching. Numbers denote vertex ids; colors represent labels. 
       Pattern $p$ is a template graph. 
       Graph $a$, $b$ and $c$ are instances of pattern $p$ in the input graph.
       These instances are called embeddings.
       Isomorphic embeddings $a$ and $b$ have same pattern $p$.
       Same embedding $b$ and $c$ also are called automorphic.  }
       \label{fig:isoauto}
\end{figure}

However, the TLV programming paradigm cannot readily formulate graph mining problems. 
Given an input graph, graph mining problems often require exploring a very large number of subgraphs and finding patterns that match some interesting criteria desired by the user.
In this paper, we use \textit{pattern} and \textit{embedding} to denote two types of subgraphs in an input graph.
A \textit{pattern} is a template, while an \textit{embedding} is an instance.
We denote $k$-embedding for an embedding contains $k$ vertices.
Embeddings are \textit{isomorphic} if they contain different vertices and edges but they have the same pattern.
Figure \ref{fig:isoauto} illustrates an example of pattern matching, which is also a step of the frequent subgraph mining.
To identify which pattern is frequent in an input graph, we should explore all embeddings, then patternize each embedding and statistic all patterns.
The exploration of subgraphs can be executed as \textit{vertex-induced} and \textit{edge-induced}.
A \textit{vertex-induced} exploration expands one vertex to an embedding in each iteration, while an \textit{edge-induced} exploration expands one edge.

The first challenge in graph mining applications is how to build a compact data structure for the intermediate data (embeddings) and process them efficiently.
In querying a $k$-vertex-pattern in a graph with $N$ distinct vertices, the time complexity is $O(N\cdot \bar{d}^{k-1})$, in which $\bar{d}$ is the average degree of the graph \cite{aggarwal:managing}. Generally, we have up to $O(N\cdot \bar{d}^{k-1})$ different embeddings of size $k$ in this graph. 
For example, the exploration of $4$-embeddings over Patent (3.8 M vertices, 16.5 M edges) \cite{leskovec:graphs} produces 13.5 billion embeddings.
The second challenge is how to efficiently test embedding isomorphism in computing patterns.
The graph isomorphism problem (GI) is an NP hard problem and no polynomial time algorithm is known \cite{shang:taming}. The fastest proven running time for GI has stood at $e^{O(\sqrt{n \log n})}$ \cite{babai:computational}.

\subsection{Limitations of State-of-the-Art Systems}

Recent graph mining systems use declarative models to solve mining problems. Arabesque \cite{teixeira:arabesque} proposes a natural programming paradigm, ``think like an embedding (TLE)'', which is also called \textit{subgraph-centric} model. 
RStream\cite{wang:rstream} employs a GRAS programming model that uses a combination of ``gather-apply-scatter'' (GAS) and relational algebra to support mining algorithms.


Arabesque is a Giraph-based distributed graph mining system.
Arabesque designs a prefix-tree-liked embeddings data structure. 
It stores $k$ arrays for $k$-embeddings, in which the $i^{th}$ array contains the ids of all vertices in the $i^{th}$ position in any embedding.
Vertex $v$ in the $i^{th}$ array is connected to vertex $u$ in the $(i+1)^{th}$ array if there exists at least one canonical embedding with $v$ and $u$ in position $i$ and $i+1$ respectively in the original set.
However, in the pattern aggregation phase, an extra canonically checking for each embedding is inevitable.
For the experiment of Arabesque, the extra checking still accounts for around 5\% of the run-time in mining applications.

RStream is a single-machine graph mining system based on X-Stream\cite{roy:xstream}.
It only supports the edge-induced embedding exploration.
When solving some vertex-based applications, like the motif counting and the clique discovery, it needs more iterations and more disk I/O.
For example, to find $4$-motifs in an input graph, RStream iterates 6 times to explore all kinds of $4$-motifs ($\binom{4}{2}=6$).
To explore all possible embeddings, RStream executes the operation of all-join in the relational algebra, which produces a huge amount of intermediate data.
For example, running the $4$-motifs counting on RStream over MiCo (100 K vertices, 1.1 M edges) \cite{elseidy:grami} produces around 1.64 TB intermediate data, while the amount of $4$-motifs in MiCo is around 11 billion.

Both Arabesque and RStream use a graph library, bliss\cite{junttila:bliss}, which is an open source tool for computing graph isomorphism problems. Bliss builds a search tree and calculates a hash value for each pattern. If two patterns contain the same hash value, they are automorphic. 
However, building the search tree brings frequently memory allocating and deallocating which slow down the processing of hashing patterns and consumes a huge amount of memory.
For example, the overhead of allocation and deallocation are more then 53\% in running $3$-FSM over Patent graph (37 labels) with support 1 and it consumes 16.1 GB memory for total 25,083 patterns.

\subsection{Our Approaches}

\begin{figure}
    \centering
    \subfloat[Adjacency matrix\label{fig:cube-a}]{
        \includegraphics[width=0.45\linewidth]{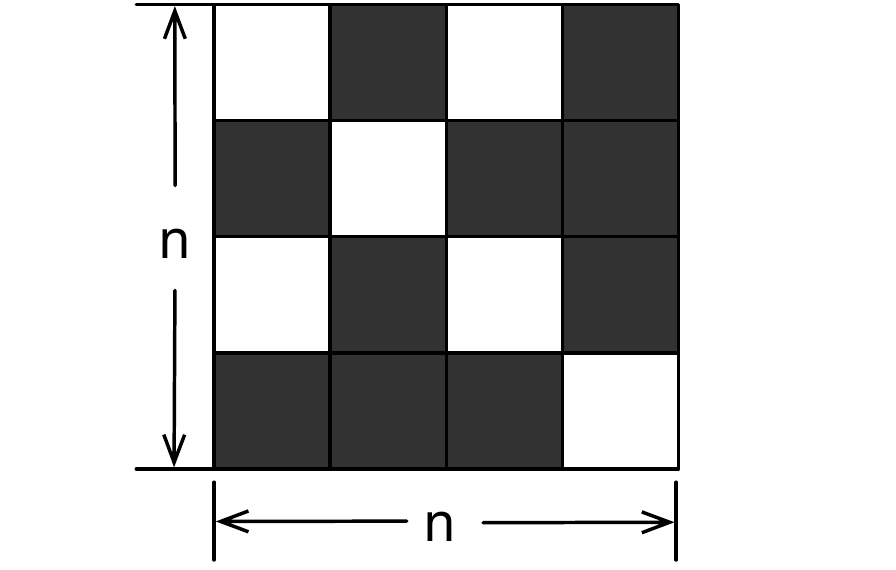}
    }
    \hfill
    \subfloat[$3$-Embedding cube\label{fig:cube-b}]{
        \includegraphics[width=0.45\linewidth]{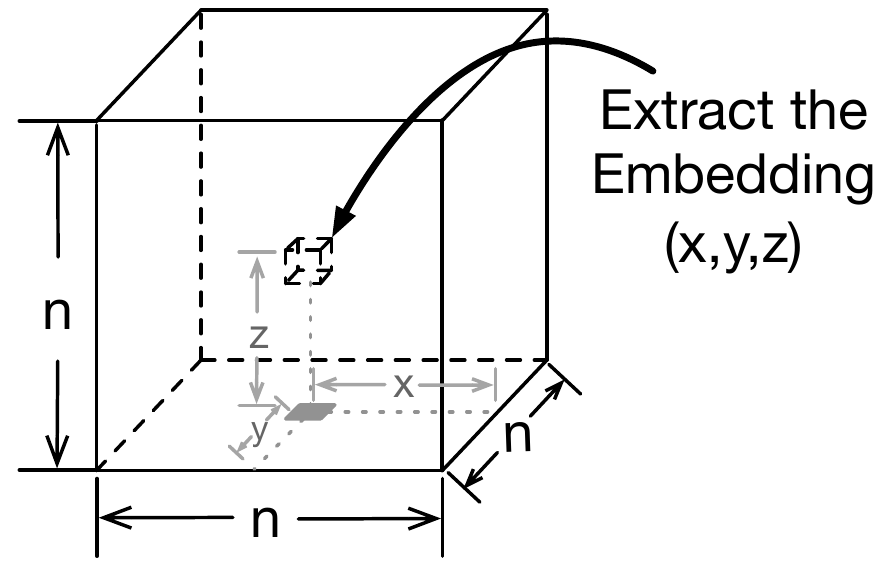}
    }
    \caption{Adjacency matrix and the $3$-embedding cube. The black blocks in Figure \ref{fig:cube-a} indicate edges in the graph. Figure \ref{fig:cube-b} indicates a cube (tensor) of $3$-embeddings and an operation of extracting an arbitrary embedding.}
    \label{fig:cube}
\end{figure}

To address the limitations of existing systems, we propose Kaleido, a single-machine, out-of-core graph mining system. 
Inspired by Arabesque, Kaleido adopts the embedding-centric computation model and presents a general programming API which fits most of graph mining applications. 
Intuitively, the set of $1$-embeddings is the vertex set of the input graph; the set of $2$-embeddings is the edge set without duplicated edges of the input graph, which can be represented by an adjacency matrix (see Figure \ref{fig:cube-a}); the set of $3$-embeddings is the set of $3$-chains and triangles, which can be represented by a cube (see Figure \ref{fig:cube-b}). 
In other words, each vertex-induced expanding of embeddings is equivalent to ascending a dimension for the intermediate data.
Thus Kaleido treats the intermediate data of the $i^{th}$ exploration as an $i$-dimension tensor. 
Inspired by compressed sparse column (CSC) for sparse matrices \cite{saad:csr}, we design a level-by-level succinct data structure of intermediate embeddings, which is called \textit{compressed sparse embedding} (CSE). 
Each iteration of the exploration ascends a dimension for the intermediate data and expands a level in CSE.

Calculating large embeddings of graph mining problems is hard to implement because of the exponential growth of the intermediate data in graph mining problems. 
For example, the sizes of embeddings in 7 iterations of the embedding explorations over a small graph CiteSeer (3.3 K vertices, 4.7 K edges) are 3.3 K, 4.5 K, 24.5 K, 352.2 K, 7.7 M, 168.2 M, 3.5 G respectively.
Kaleido focuses on efficiently solving the GI problems in small embeddings (less than 9 vertices).
From the property of isomorphism, we explain that isomorphic embeddings have the same eigenvalues.
Harary \emph{et at.} \cite{harary:cospectral} proved that the smallest non-isomorphic unlabeled graphs contain 6 vertices with the same eigenvalues and the smallest non-isomorphic unlabeled graphs with the same eigenvalues and the same vertex degrees contain 9 vertices.
To solve the labeled graph isomorphism problem, Kaleido combines the label information and the degree information of vertices in each pattern and eigenvalues of the adjacency matrix to check isomorphism over embeddings.

To store more intermediate data when the scale of the input graph or the exploration depth increases, we design a hybrid storage for the intermediate embedding.
According to the level-by-level structure of CSE, when the memory is insufficient to afford the whole intermediate data, the hybrid storage stores large levels of CSE on disk.
To balance the work load in processing the intermediate data on disk, Kaleido predicts the capacity of embeddings candidate in the next iteration, then divides the exploring tasks to each thread evenly according to the prediction.
When processing the intermediate data on disk, Kaleido adopts a slide-window strategy to guarantee the performance of parallel computation and hide the overhead of I/O.

To summarize, we make the following contributions:

\begin{itemize}
       \item We introduce the computation model of Kaleido in Section \ref{sec:design}. We propose a novel succinct embeddings data structure, compressed sparse embedding, which treats embeddings as a sparse tensor. 
       \item We propose a lightweight graph isomorphism checking algorithm in using eigenvalues of the adjacency matrix of each pattern (Section \ref{sec:design}).
       \item We implement a hybrid storage for the large intermediate data, which treats disks as an extension of the memory and hides the overhead of I/O by the computation of the embedding exploration and pattern aggregation (Section \ref{sec:impl}).
       \item We design an API for popular graph mining applications, which enables embedding exploration and pattern aggregation to be expressed effectively. We present four popular graph mining applications which are expressed in Kaleido API (Section \ref{sec:api}).
       \item We compare Kaleido with Arabesque and RStream in four graph mining applications. We compare the performance of GI problems in Kaleido with bliss. We demonstrate the scalability of Kaleido in different applications. We demonstrate the I/O performance in the hybrid storage (Section \ref{sec:evl}).
\end{itemize}

The rest of paper are organized as follows. Section \ref{sec:pre} formalizes the problem. Section \ref{sec:design} presents the computation model of Kaleido, the design of CSE and the eigenvalue-based isomorphism checking algorithm. Section \ref{sec:impl} presents the storage strategy of the large intermediate data. Section \ref{sec:api} introduce the API of Kaleido and the implementations of popular graph mining applications. Section \ref{sec:evl} presents the experimental evaluation. Section \ref{sec:rw} surveys related works and Section \ref{sec:conclusion} concludes.

\section{Preliminaries}\label{sec:pre}

A \textit{graph} $G=(V,E,L)$ consists of a set of vertices $V$, a set of edges $E$ and a labeling function $L$ that assigns labels to vertices and edges.
A graph $G'=(V',E',L')$ is a \textit{subgraph} of graph $G=(V,E,L)$, i.e., $V'\subseteq V$, $E'\subseteq E$ and $L'(v)=L(v),\forall v\in V'$.
A \textit{pattern} is a template graph, while an \textit{embedding} is an instance. 
In this paper, the vertex-induced embedding is noted as $e=\langle v_1,...,v_k\rangle$.
If an embedding contains $k$ vertices, we say that the size of embedding $e$ is $k$. 
The edge-induced embedding is analogous.
 
\begin{definition}
       We say that a subgraph $G_a=(V_a,E_a,L_a)$ of graph $G$ is \textit{isomorphic} to another subgraph $G_b=(V_b,E_b,L_b)$ of $G$ if and only if there exists a bijection $f_{ab}$ between $G_a$ and $G_b$, such that (i) $L_a(v)=L_b(f_{ab}(v)),\forall v\in V_a$, and (ii) $(f_{ab}(u),f_{ab}(v))\in E_b$ and $L_a(u,v)=L_b(f_{ab}(u),$ $f_{ab}(v)),$ $\forall (u,v)\in E_a$.
       \label{def:ssiso}
\end{definition}

In Figure \ref{fig:isoauto}, both subgraphs $a=(V_a,E_a,L_a)$ and $b=(V_b,E_b,L_b)$ have pattern $p$. There exists a bijection $f$ between subgraphs $a$ and $b$, $f_{a\leftrightarrow b}:\{1\leftrightarrow 3, 2\leftrightarrow 2, 5\leftrightarrow 5\}$, which satisfies constraining of subgraphs isomorphism. Thus subgraph $a$ is isomorphic to subgraph $b$.
Two subgraphs are \textit{automorphic} if and only if they contain the same edges and vertices. As shown by Figure \ref{fig:isoauto}, subgraphs $b$ and $c$ contain the same edges and vertices, thus they are automorphic. 


\begin{definition}
       We say an embedding $e=\langle v_1,...,v_n\rangle$ of graph $G=(V,E)$ is \textit{canonical} if (i) $\forall i>1$ it holds $v_i>v_1$; (ii) $\forall i>1, \exists j<i$ satisfies that $(v_j,v_i)\in E$; (iii) $\forall v_a, v_b, v_c$ if $a<b<c$, $(v_a,v_c)\in E$ and $\nexists d<a$ satisfies that $(v_d,v_c)\in E$, it holds $v_b<v_c$.
       \label{def:can}
\end{definition}

In other words, if an embedding is canonical, it should hold the following three properties. (i) The id of the first vertex in the embedding is the minimum value. (ii) Each vertex in the embedding must be a neighbor of the vertex which is indexed a smaller id, except the first vertex. (iii) There exists an edge in the embedding, $(v_a,v_c),a<c$ and all of vertices before $v_a$ are not neighbor of $v_c$, therefore if any vertex exists between $v_a$ and $v_c$, it must holds that $v_b<v_c$. 


\section{Computation Model}\label{sec:design}

In this section, we describe the computation model of Kaleido and the design of data structure of intermediate data and patterns.
The procedure of graph mining applications in Kaleido are mainly divided into two phases, the embedding exploration phase and the pattern aggregation phase.

\subsection{Embedding Exploration}\label{subsec:exp}

In the phase of embedding generation, given the size of embedding, our goal is to generate all of the possible and unique embeddings. Then according to the user's criteria, we eliminate embeddings which are ineligible. We introduce the \textit{canonical filter} which guarantees that embeddings are complete and unique. 

%
%
%
%

We follow Arabesque's idea of checking embedding canonicality for each candidate.
From Definition \ref{def:can}, an embedding $e$ is canonical if and only if its vertices were visited in the following order: start by visiting the vertex with the smallest id and then recursively add the neighbor of $e$ with the smallest id that has not been visited yet.

\begin{figure}
       \centering
       \includegraphics[width=0.5\textwidth]{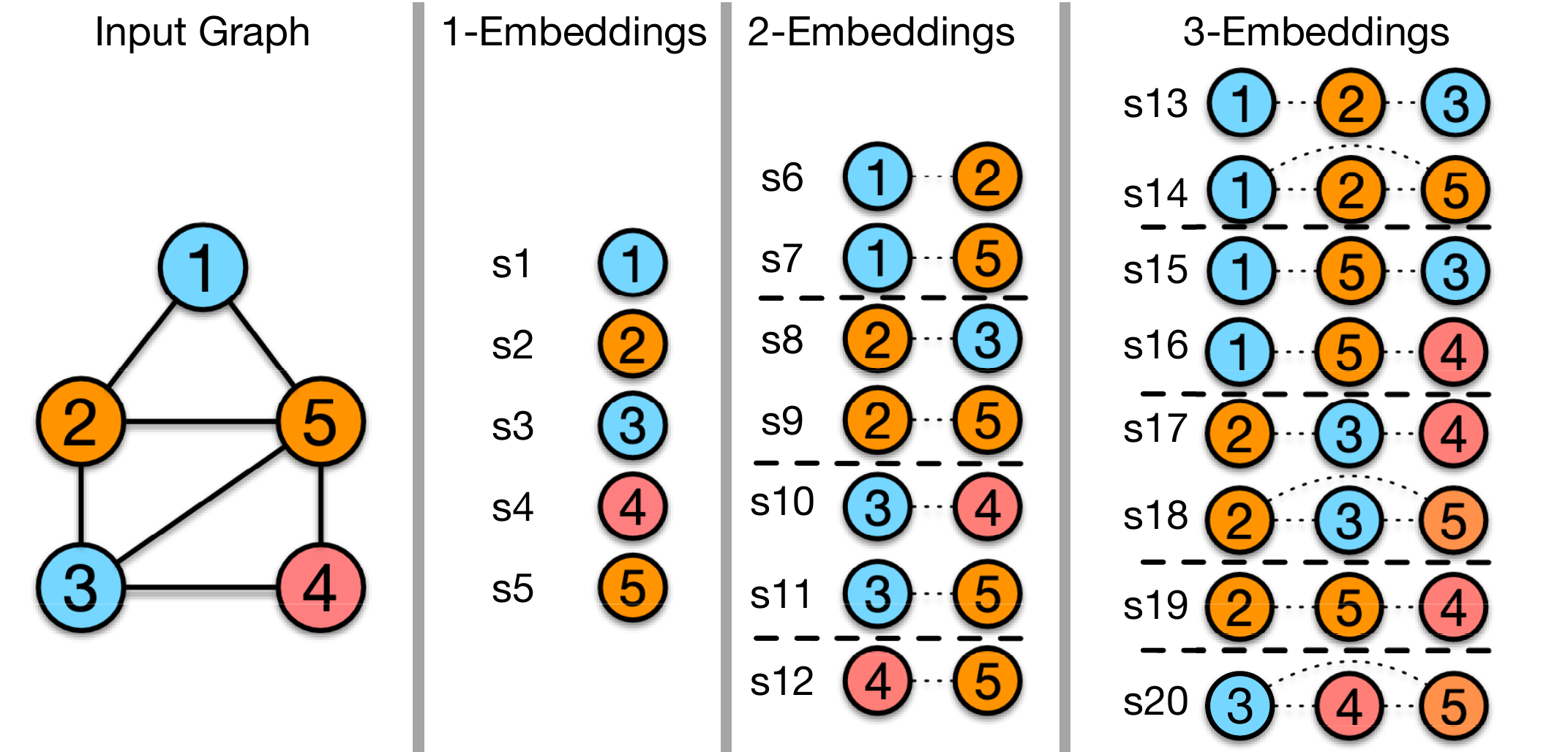}
       \caption{Procedure of vertex-based generating canonical $3$-embeddings. $2$-embeddings and $3$-embeddings in the figure are arrays of vertices ids, and the order of each embedding is immutable; the dotted lines between two embeddings, like a dotted line between $s_7$ and $s_8$, which means $s_6$ and $s_7$ are generated by $s_1$ and $s_8$ and $s_9$ are generated by $s_2$; note that the dotted lines between vertices do not really exist in the embedding array, instead they only represent there exist an edge in input graph between these two vertices.}
       \label{fig:embgen}
\end{figure}

Figure \ref{fig:embgen} illustrates the process of a series of vertex-based explorations to $3$-embeddings.
Without loss of generality, consider an exploration of a $2$-embedding $s_8=\langle 2,3\rangle$.
First, neighbors or candidates of $s_8$ in $G$ are $\{1,4,5\}$, thus possible $3$-embeddings generated by $s_8$ are $\langle 2,3,1\rangle$, $\langle 2,3,4\rangle$ and $\langle 2,3,5\rangle$.
This step guarantees completeness of this exploration.
Next, according to Definition \ref{def:can}, $\langle 2,3,1\rangle$ does not satisfies property (i) of canonical embedding, for $1<2$ and $2$ is the first vertex of $s_8$, while both $\langle 2,3,4\rangle$ and $\langle 2,3,5\rangle$ are canonical.
Therefore $s_{17}$ and $s_{18}$ are generated.

\subsubsection{Embeddings Data Structure}\label{subsec:embmem}

The goal of storing embeddings is divided into two parts: (i) minimizing memory usage and (ii) obtaining an arbitrary embedding as fast as possible.
Like an adjacency matrix form of a graph, a $k$-embedding set can be treated as an adjacency $k$-dimension tensor (see Figure \ref{fig:cube}).
Kaleido stores the graph structure in \textit{compressed sparse column (CSC)}, which is equivalent to the sparse adjacency matrix of the graph. 
Inspired by CSC, we design a succinct data structure for embeddings, which is called \textit{compressed sparse embedding (CSE)}. If a $k$-embedding set is stored in CSE, we call it $k$-CSE.

\begin{figure}[h]
    \centering
    \subfloat[$2$-Embeddings\label{fig:csv-a}]{
        \includegraphics[width=0.47\linewidth]{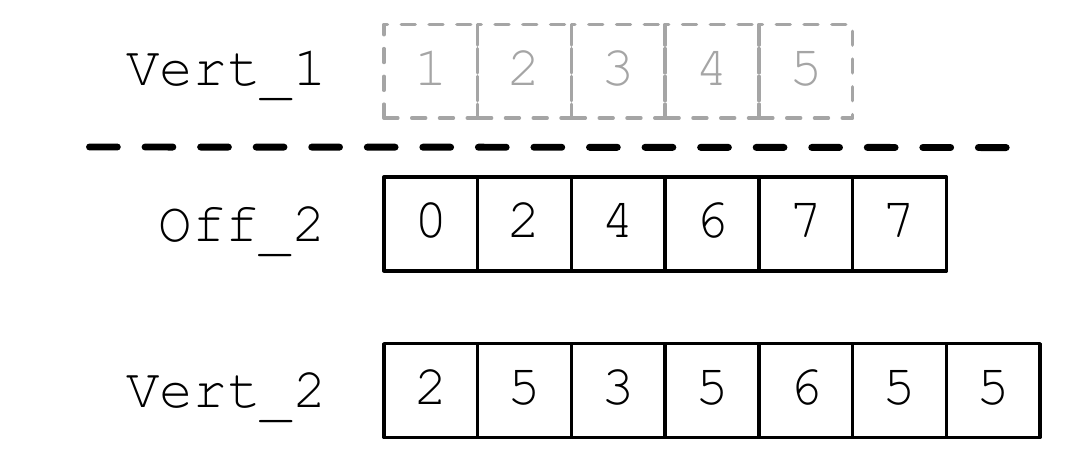}
    }
    \hfill
    \subfloat[$3$-Embeddings\label{fig:csv-b}]{
        \includegraphics[width=0.47\linewidth]{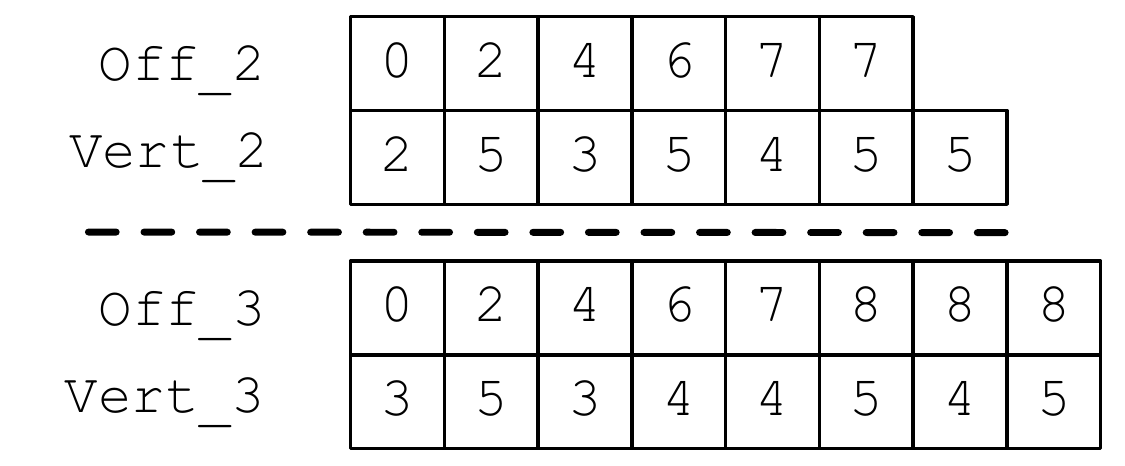}
    }


    \caption{The structure of compressed sparse embedding (CSE). This figure shows a $2$-CSE and a $3$-CSE of $2$-embeddings and $3$-embeddings illustrated in Figure \ref{fig:embgen} respectively. The gray array $vert_1$ does not really store in Kaleido; it indicates the relationship between vertex array and offset array. The dotted lines divide CSE into different levels.}
    \label{fig:csv}
\end{figure}

As illustrated in Figure \ref{fig:csv}, Kaleido stores embeddings level-by-level.
In each level, the structure of embeddings is stored in two arrays.
Vertex array ($vert_l$) indicates the last vertex of each embedding in level $l$.
Offset array ($off_l$) indicates the start offset $off_l(i)$ and end offset $off_l(i+1)$ in vertex array of level $l$.
A slice of vertex array, $\big[off_l(i),off_l(i+1)\big)$, indicates that these vertices possesses same embedding prefix. 
For example, in Figure \ref{fig:csv-a}, the first two elements of offset array are $0$ and $2$ which indicates a slice of vertex array $\{vert_2(i)|i\geq 0, i < 2\}=\{2,5\}$.
It correspond to $s_6$ and $s_7$ illustrated in Figure \ref{fig:embgen}, which possess the same embedding prefix $\{1\}$.
Therefore each vertex in vertex array corresponds to a unique embedding in current level and an embedding prefix of next level.
Therefore the length of vertex array in level $i$ is equal to the length of offset array in level $i+1$ minus $1$ (to compute conveniently, last element indicate the length of $vert_i$).

Now given an arbitrary offset of vertex array in level $k$, we can obtain the $k$-embedding corresponding to this offset.
For example, given offset $5$ of vertex array in level $3$ in Figure \ref{fig:csv-b}, the goal is to find the corresponded $3$-embedding.
First, we note the last element of this embedding is $vert_3(5)=5$, $\langle \cdot,\cdot,5\rangle$.
Then we find that offset $5$ is greater than $off_3(2)=4$ and less than $off_3(3)=6$, thus the coordinate of offset $5$ in offset array in level $3$ is $2$.
Next, we do this processing again in level $2$, and the offset of the vertex array is $2$.
At last, we obtain the $3$-embedding $\langle 2,3,5\rangle$, which corresponds to $s_{18}$ in Figure \ref{fig:embgen}.

\textbf{Complexity}: Each iteration of the embedding exploration extends $O(\bar{d})$ space ($\bar{d}$ is the average of vertex degree and $\bar{d}\propto|E|/|V|$). Thus the space complexity of $k$-CSE is $O(|E|^{k-1}/|V|^{k-2})$. Given an arbitrary offset of vertex array in level $k$, the time complexity of obtain the corresponding embedding is $O(k\log\bar{d})=O(\log(|E|/|V|))$.

\subsection{Pattern Aggregation}\label{subsec:iso}

After $k$ iterations of the embedding exploration, Kaleido collects all possible canonical embeddings in the input graph, whose size is no more than $k$.
Then int the pattern aggreation phase, Kaleido calculates the pattern of each embedding and aggregates them.
The challenge of the aggregation is how to efficiently test embedding isomorphism and obtain the pattern.
As it is well studied, GI is known as an NP-hard problem \cite{shang:taming}.
The state-of-art algorithm is to solve the problem has run-time $e^{O(\sqrt{n\log n})}$ for graphs with $n$ vertices \cite{babai:computational}.
Existing algorithms or libraries build search tree for each pattern to solve the GI problem, like Bliss.

\begin{figure}
    \centering
    \subfloat[Input Graph\label{fig:ptnst-a}]{
        \includegraphics[width=0.25\linewidth]{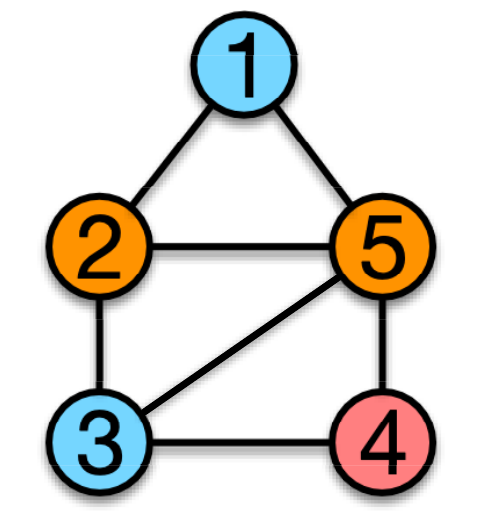}
    }
    \hfill
    \subfloat[Adj. Matrix\label{fig:ptnst-b}]{
        \includegraphics[width=0.25\linewidth]{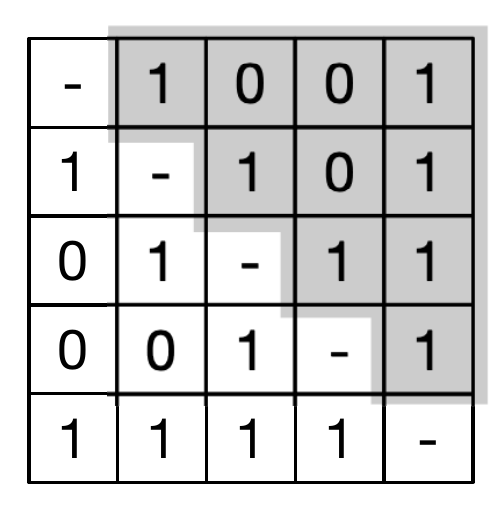}
    }
    \hfill
    \subfloat[Pattern Structure\label{fig:ptnst-c}]{
        \includegraphics[width=0.4\linewidth]{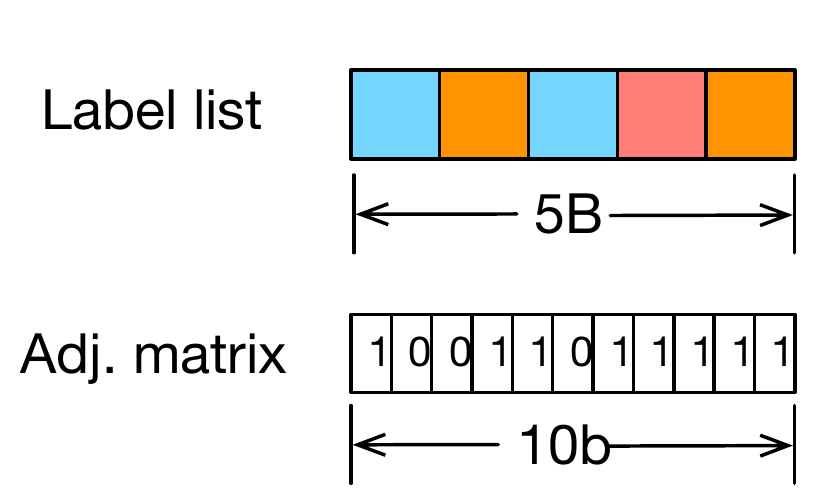}
    }
    \caption{Pattern structure. Figure \ref{fig:ptnst-a} indicates an undirected graph. Figure \ref{fig:ptnst-b} is adjacency matrix of the graph. Figure \ref{fig:ptnst-c} consists 2 parts of pattern structure. Colors in label list indicate different labels in the graph. Adjacency matrix is stored in form of bitmap.}
    \label{fig:ptnst}
\end{figure}

In Kaleido, patterns are stored in a simple compact data structure, as illustrated in Figure \ref{fig:ptnst}.
The data structure contains each pattern's label information and structural information.
Generally, we use an adjacency matrix to indicate the structural information of this pattern and a label array to indicate vertex labels.
The order of labels matches with an adjacency matrix.
Kaleido stores the up-triangle part of adjacency matrix (gray area in Figure \ref{fig:ptnst-b}) in form of $1$-dimension array and stores it as a bitmap.
Obviously, storing an $k$-pattern in this data structure needs a label array whose size is $k$, and a bitmap whose size is $\frac{1}{2}(k(k-1))$.


In pattern aggregation phase, Kaleido directly transforms each embedding to this pattern data structure.
Then Kaleido should identify non-isomorphic patterns and automorphic patterns.
Note that the relationship between the pattern and the data structure is 1 to n.
One data structure represents a unique pattern but one pattern can be represented in different data structures.
These different data structures represent automorphic patterns.
In Kaleido, the GI problem is solved by utilizing the relationship between the adjacency matrix of the pattern and its eigenvalues.

\begin{theorem}\label{theo:eig}
       Let $G_a$ and $G_b$ be two $k$-subgraphs of an undirected graph $G$, $A$ and $B$ be adjacency matrices of $G_a$ and $G_b$ respectively, $\Lambda_a=\{\lambda_{a1},...,$ $\lambda_{an}\}$ and $\Lambda_b=\{\lambda_{b1},...,\lambda_{bn}\}$ be eigenvalues of $A$ and $B$ respectively. If $G_a$ is isomorphic to $G_b$, it holds that $\Lambda_a=\Lambda_b$.
\end{theorem}

\begin{proof}
       Let $G_{pa}$ and $G_{pb}$ be patterns of $G_a$ and $G_b$ respectively. $A$ and $B$ are also adjacency matrices of $G_{pa}$ and $G_{pb}$ respectively. According to Definition \ref{def:ssiso}, if $G_a$ is isomorphic to $G_b$, $G_{pa}$ is automorphic to $G_{pb}$. It leads that $A$ can be transformed to $B$ by a similarity transformation. Thus there exists an invertible n-by-n matrix $P$, such that
       $$B=P^{-1}AP$$
       Thus $A$ and $B$ are similar. Similar matrices have the same eigenvalues and their algebraic multiplicities are the same.
\end{proof}

\begin{figure}[]
    \centering
       \includegraphics[width=0.5\textwidth]{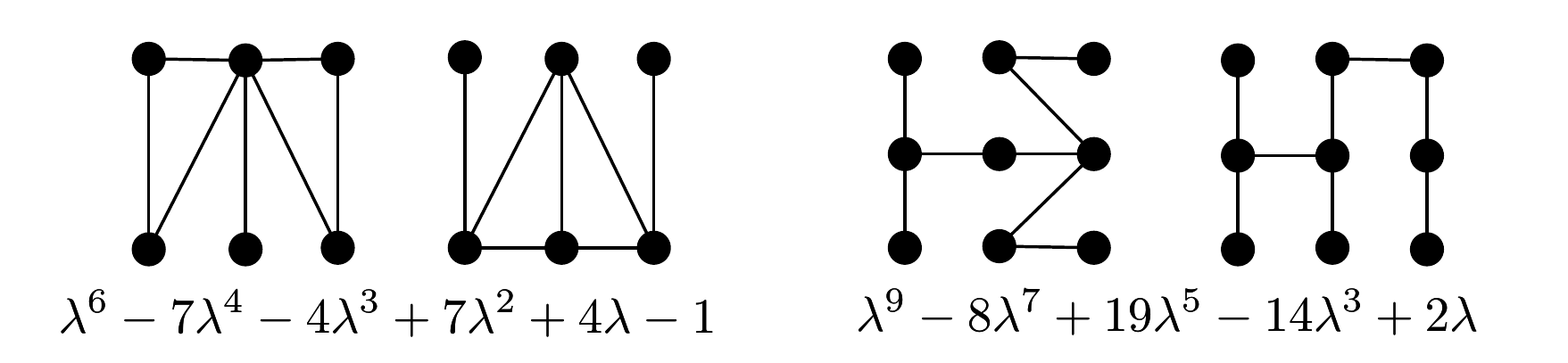}
    \caption{The smallest counterexamples and corresponding characteristic polynomials.}
    \label{fig:counterexample}
\end{figure}

According to Theorem \ref{theo:eig}, if eigenvalues of two patterns are different, it holds that these patterns are non-isomorphic.
Unfortunately, not all graphs that have the same eigenvalues \footnote{In this paper, if graphs have the same eigenvalues, their algebraic multiplicities are the same as well.} are necessarily isomorphic.
However, Harary \emph{et at.} \cite{harary:cospectral} proved that the smallest non-isomorphic graphs with the same eigenvalues contain 6 vertices and the smallest non-isomorphic graphs with the same eigenvalues and the same vertex degrees contain 9 vertices (see Figure \ref{fig:counterexample}).
Thus we get a corollary.

\begin{corollary}\label{cor:iso}
       If two $k$-embeddings ($k<6$) have the same eigenvalues, they are isomorphic.
       Further, if two $k$-embeddings ($k<9$) have the same vertex degrees and the same eigenvalues, they are isomorphic.
\end{corollary}

\begin{algorithm}[h]
       \footnotesize
       \caption{Kaleido's graph isomorphic check \label{alg:iso}}
       \SetKwInput{In}{Input}
       \SetKwInput{Out}{Output}
       \SetKwProg{Fn}{Func}{\string:}{end}
       \SetKwFunction{Init}{Init}%
       \SetKwFunction{Swap}{Swap}%
       \SetKwFunction{WAMatrix}{WeightedAdjMatrix}
       \SetKwFunction{CharPoly}{CharPloynomical}%
       \SetKwFunction{EigenHash}{EigenHash}%
       \In{Embedding $e=\langle v_1,...,v_k\rangle$}
       \Out{Hash value $hash\_value$ of Embedding $e$}
       \Fn{\Init{$e$}}{
              Label array $L \gets \{l_i=label(v_i) | v_i \in e,\forall i\in [1,k] \}$ \\
              Adjacency matrix $\mathbf{A} \gets \{a_{i,j}=CheckLink(v_i,v_j) | v_i, v_j \in e, i<j, \forall i,j \in [1,k] \}$\\
              Degree array $D \gets \{d_i=\deg(v_i) | v_i \in e,\forall i\in [1,k] \}$ \\
              \Return{$L, \mathbf{A}, D$}
       }
       \Fn{\Swap{$i,j$}}{
              Swap $l_i$ and $l_j$\\
              \For{$1 \leq t \leq k$}{
                     Swap $a_{i,t}$ and $a_{j,t}$\\
                     Swap $a_{t,i}$ and $a_{t,j}$
              }
              Swap $d_i$ and $d_j$
       }
       \Fn{\WAMatrix{$L, \mathbf{A}$}}{
              $\mathbf{M} \gets \{m_{i,j}=0|\forall i,j \in [1,k]\}$\\
              \For{$1 \leq i < j \leq n $}{
                     \If{$a_{i,j}=1$}{
                            $m_{i,j} \gets l_i | l_j$ \\
                            $m_{j,i} \gets l_i | l_j$
                     }
              }
              \Return{$\mathbf{M}$}
       }
       \Fn{\CharPoly{$\mathbf{M}$}}{
              Characteristic polynomial $P \gets \{p_i=0|\forall i \in [1,k] \}$\\
              $\mathbf{C} \gets \mathbf{M}$\\
              \For{$1 \leq i \leq k$}{
                     \If{$i > 1$}{
                            $\mathbf{C} \gets \mathbf{M} \cdot (\mathbf{C}+p_{k-i+1}\mathbf{I}_k)$
                     }
                     $p_{i-k}=-\frac{tr(\mathbf{C})}{k}$
              }
              \Return{$P$}
       }
       \Fn{\EigenHash{$e$}}{
              $L,\mathbf{A},D \gets$ \Init{$e$}\\
              \For{$1 \leq i < j \leq n $}{
                     \uIf{$l_i > l_j$}{
                            \Swap{$i,j$}
                     }
                     \ElseIf{$l_i = l_j$ and $d_i>d_j$}{
                            \Swap{$i,j$}
                     }      
              }
              $\mathbf{M} \gets$ \WAMatrix{$L, \mathbf{A}$}\\
              $P \gets$ \CharPoly{$\mathbf{M}$}\\
              $hash\_value \gets hash(L)\oplus hash(D)\oplus hash(P)$\\
              \Return{$hash\_value$}
       }
\end{algorithm}

When vertices of the input graph have different labels, the GI problem becomes a little complex.
Algorithm \ref{alg:iso} illustrates the solution of the GI problem in Kaleido where the size of embedding is less than 9.
Kaleido maintains the vertex label array $L$ in an ascending order (lines 30-31) and the degrees ($D$) of the same label vertices in an ascending order as well (lines 32-33).
Note that \texttt{Swap} function also maintains the adjacency matrix $A$, so that the vertex order in $A$ is consisting with $L$ and $D$.
Then Kaleido builds a weighted adjacency matrix $M$ (line 34, lines 12-18) whose edge weights is a concatenation of two vertex labels (lines 16-17). 
Note that label $l_i$ is no more than label $l_j$ after the sorting in lines 29-33;
Next, Kaleido calculates the eigenvalues of the matrix $M$.
However, the calculation of the accurate eigenvalues is redundant, while Kaleido calculate the characteristic polynomial of the matrix $M$ by the Faddeev-LeVerrier algorithm (line35, lines 19-26).
Finally, Kaleido calculates the hash value of each embedding by combining the hash value of the label array $L$, the degree arrary $D$ and the characteristic polynomial $P$ in XOR ($\oplus$).

\begin{theorem}
       \label{theo:alg}
       Let $e_1$ and $e_2$ be two $k$-embeddings of an undirected graph $G$, $k < 9$.
       Let $h_1$ and $h_2$ be hash values of $e_1$ and $e_2$, which are calculated by Algorithm \ref{alg:iso}.
       Embedding $e_1$ is isomorphic to embedding $e_2$ if and only if $h_1 = h_2$.
\end{theorem}

\begin{proof}
       For embeddings $e_1$ and $e_2$, hash values $h_1 = h_2$ is equivalent to that label arrays $L_1 = L_2$, degree arrays $D_1 = D_2$ and characteristic polynomials $P_1 = P_2$.
       From Definition \ref{def:ssiso}, the isomorphism leads to $L_1=L_2$ and $D_1=D_2$.
       Theorem \ref{theo:eig} satisfies $P_1=P_2$.
       The necessity is proved.
       Note $M_1, M_2$ as the weighted adjacency matrix of $e_1$ and $e_2$ respectively.
       Note $\epsilon_1$ as a graph which is no vertex label but the adjacency matrix is weighted and equals to $M_1$.
       Note $\epsilon_2$ symmetrically.
       From Corollary \ref{cor:iso}, $D_1 = D_2$ and $P_1 = P_2$ guarantee that $\epsilon_1$ and $\epsilon_2$ are isomorphic. 
       Thus, there exists a bijection $f(u)=v,\forall u\in V_{\epsilon_1}, \forall v\in V_{\epsilon_2}$ between $V_{\epsilon_1}$ and $V_{\epsilon_2}$, such that $(f(u),f(v))\in E_{\epsilon}$ and $L(u,v)=L(f(u),f(v)),\forall (u,v)\in E_{\epsilon_1}$.
       Then combining $L_1=L_2$ and $D_1=D_2$, it leads to that $L(u)=L(f(u)), \forall u\in V_{e_1}$.
       The sufficiency is proved.
\end{proof}

\section{Implementation}\label{sec:impl}

In this section, we introduce the storage strategy of larger intermediate data in Kaleido.
Then we introduce the load-balance strategy of Kaleido in facing the insufficient RAM.

\subsection{Embedding Hybrid Storage}\label{subsec:hybrid}

According to the space complexity of CSE, exploring $(k+1)$-embeddings from $k$-CSE needs extra $O(|E|^{k}/|V|^{k-1})$ space.
The memory would be insufficient when the exploration depth increases.
Thanks to the level-by-level structure of CSE, Kaleido stores large levels of CSE on disk intuitively.
We call this half-memory-half-disk storage the hybrid storage.
\begin{figure}
       \centering
           \includegraphics[width=0.5\textwidth]{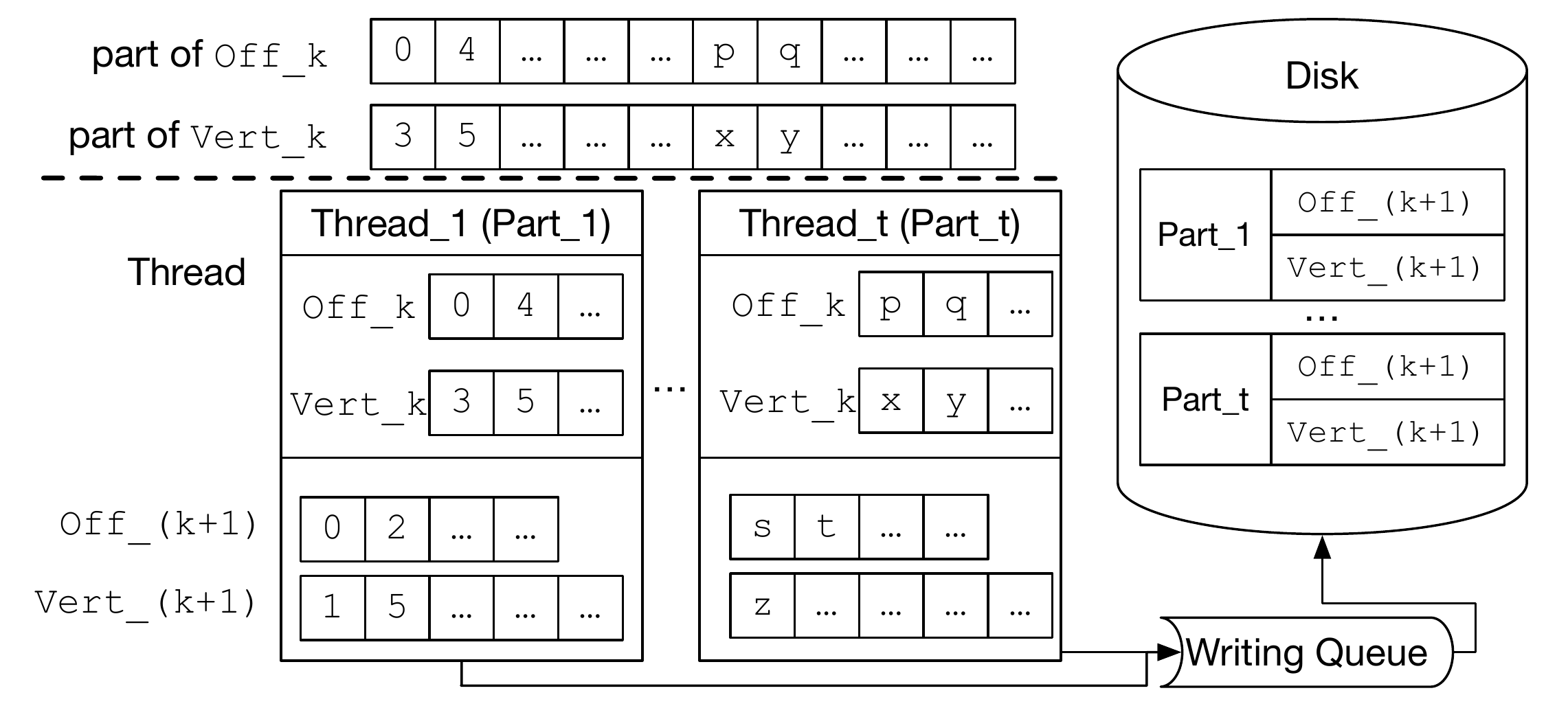}

       \caption{Exploration on Hybrid CSE. The first $k$ levels are stored in memory. The $(k+1)^{th}$ level is stored on disk in $t$ parts (in this example, $t$ equals to the thread number).} 
       \label{fig:hybrid}
\end{figure}

First, Kaleido partitions $vert_k$ into several parts continuously and evenly and assigns to each thread.
Then each thread calculates the $(k+1)^{th}$ elements of each $k$-embedding and records the offset when all canonical candidates of an $k$-embedding are enumerated.
Finally, each thread appends their part of $vert_{k+1}$ to the writing queue and the writing queue flushes these parts to disk (see Figure \ref{fig:hybrid}).
If memory is sufficient, Kaleido merges $t$ parts of $off_{k+1}$ in memory, otherwise appends each part of $off_{k+1}$ to the writing queue.
When Kaleido explores $(k+2)$-embeddings and constructs the $(k+2)^{th}$ level of embeddings, load the first part of $vert_{k+1}$ and $off_{k+1}$ (if exists) on disk.
Then Kaleido executes the former process again and stores $vert_{k+2}$ and $off_{k+2}$ to disk part by part, until it finishes the last part of $vert_{k+1}$.

To explore deeper embeddings or process embeddings in the hybrid storage, Kaleido adopts the sliding window strategy to hide the overhead of I/O.
When processing the hybrid storage embeddings, Kaleido maintains $h$ windows for $h$ levels stored on disk.
Each window respectively loads two parts of a level of CSE, which are produced by $t$ threads as shown in Figure \ref{fig:hybrid}.
When all the first parts (main part) of $h$ windows are loaded, Kaleido processes all embeddings in current windows in parallel, while the $h$ windows load the second parts (candidate part) in its corresponding level.
If the main part of a window is processed, Kaleido slides this window to the next position (swaps the main part and the candidate part, then abandons the candidate part).
Repeat this procedure until all parts on disk are processed.

\subsection{Load-balance of Hybrid Storage}

In each iteration of the embedding exploration, Kaleido expands a neighbor vertex or edge for each embedding.
Similar to the definition of the vertex degree, We say an embedding degree is the neighbors' number of the embedding.
One of the hallmark properties of natural graphs is their skewed power-law degree distribution \cite{faloutsos:power}.
The degree distribution of embeddings is also skewed power-law distribution.
When the RAM can afford the embeddings data, Kaleido utilizes a work-steal strategy to deal with the load-balance problem in the exploration.
However, when the RAM is insufficient, Kaleido stores the high level embeddings on disk in several parts.
The unbalanced partition strategy of the embedding exploration would produce huge parts which cannot load to the memory once.
The work-steal strategy can only balance the execution of the exploration but cannot balance the size of each part.

\begin{figure}
       \centering
           \includegraphics[width=0.5\textwidth]{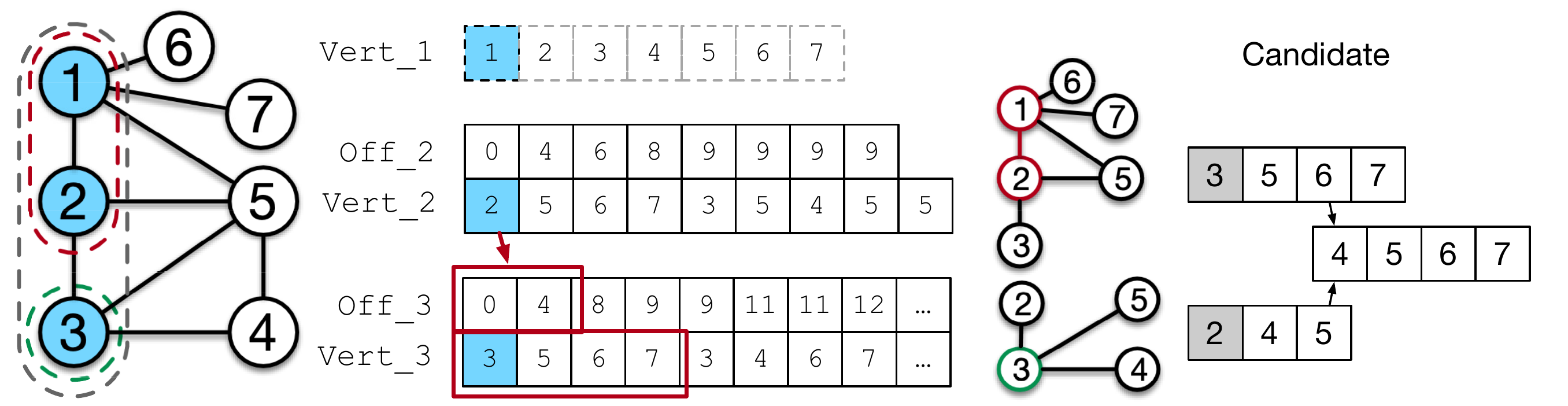}
       \caption{An example of the prediction of the candidate size of embedding $\langle 1,2,3 \rangle$.
       Candidates of $\langle 1,2,3 \rangle$ is the union of the neighbor set of $\langle 1,2 \rangle$ and the neighbor set of $\langle 3 \rangle$.} 
       \label{fig:predict}
\end{figure}

To balance the work load in the exploration of the $(k+1)^{th}$ level, Kaleido predicts the size of $vert_{k+1}$.
Figure \ref{fig:predict} illustrates an example of the prediction.
According to the structure of CSE, the neighbor set of the embedding $\langle 1,2,3 \rangle$ is the union of the neighbor set of $\langle 1,2 \rangle$ and the neighbor set of $\langle 3 \rangle$.
From offset arrays in CSE, we easily obtain the degree of $\langle 1,2 \rangle$ and $3$.
Kaleido predicts the candidate size accurately by merging the two sources of the candidate.
The time complexity of the merging is $O(\bar{d})$.
According to the prediction, Kaleido partitions the exploration tasks evenly to each threads.

\section{Kaleido API}\label{sec:api}

In this section, we demonstrate the API for graph mining problems in Kaleido.
Finally, we introduce implementations of popular graph mining applications in Kaleido.


The API of Kaleido is illustrated in Listing \ref{list:api}.
\texttt{EmbeddingFilter} and \texttt{PatternFilter} are two optional filters in the embedding exploration phase and the pattern aggregation phase respectively, which will return \texttt{true} in default.
For the vertex-induced embedding exploration, \texttt{EmbeddingFilter} works in exploring $(k+1)$-embeddings from $k$-embeddings: the \texttt{Embedding e} is a $k$-embedding and the \texttt{Vertex v} is a candidate vertex which is a neighbor of \texttt{e} normally.
It is analogous to edge-induced embedding exploration, in which the candidate of each embedding is an \texttt{Edge <u,v>}.
Actually, we eliminate a default embedding filter, the canonical embedding filter. The \texttt{vertex v} could be appended to \texttt{Embedding e} if and only if they satisfy constraints of both user-defined filter and the default canonical filter.
\texttt{Pattern Filter} works in the aggregation of patterns, in order to prune ineligible patterns.
\texttt{AggregatingMapper} and \texttt{AggregatingReducer} (\texttt{Mapper} and \texttt{Reducer} in short) must be implemented by customized applications.
\texttt{Mapper} is an operator for calculating the pattern of an embedding \texttt{e} and adding the pattern to \texttt{PatternMap}.
\texttt{Mapper} is calculated in \texttt{ResultAggregator} concurrently. \texttt{Reducer} aggregates \texttt{PatternMap}s returned by \texttt{Mapper}, and prunes patterns which are incompatible of \texttt{PatternFilter}, then returns results in form of \texttt{PatternMap}. 

\lstset{frame=tb,
  language=C++,
  aboveskip=1pt,
  belowskip=1pt,
  showstringspaces=false,
  columns=flexible,
  basicstyle={\footnotesize\ttfamily},
  numbers=none,
  numberstyle=\tiny\color{gray},
  keywordstyle=\color{blue},
  commentstyle=\color{dkgreen},
  stringstyle=\color{mauve},
  breaklines=true,
  breakatwhitespace=true,
  tabsize=3,
  caption={Kaleido API},
  label=list:api
}

\begin{lstlisting}
// Optional user defined filter functions
bool EmbeddingFilter(Embedding e, Vertex v)
bool EmbeddingFilter(Embedding e, Edge <u,v>)
bool PatternFilter(Pattern p)

// 2 functions of aggregation phases
PatternMap AggregatingMapper(Embedding e)
PatternMap AggregatingReducer(List<PatternMap> pMaps, PatternFilter pFilter)

// Main processing function in applications
List<Embedding> Init(Graph g, int depth)
List<Embedding> EmbeddingsExplorer(Graph g, List<Embedding>, EmbFilter eFilter)
PatternMap ResultAggregator(AggregatingMapper mapper, AggregatingReducer reducer)
\end{lstlisting}

The processes of graph mining application in Kaleido are described as follows. 
Initially, in the vertex-induced applications, \texttt{Init} treats the vertex set of the input graph $G$ as $1$-embeddings or the edge set as $2$-embeddings. The user should determine a terminated condition of \texttt{EmbeddingExplorer}.
While in the edge-induced application, \texttt{Init} treats the whole edge set of Graph $G$ as the set of $1$-embeddings.
Generally, the user gives the maximum number $k$ of different vertices in each embedding.
Then \texttt{EmbeddingExplorer} iterates $k-1$ times to explore all possible $k$-embeddings with restriction of \texttt{EmbeddingFilter}.
If embeddings are larger than memory, Kaleido stores them to disks intermediately.
Next \texttt{PatternComputer} calculates patterns of each $k$-embedding and aggregates patterns with restriction of \texttt{PatternFilter}.
Finally, Kaleido returns results in the form of \texttt{PatternMap}.

\subsection{Popular Mining Applications in Kaleido}


\textbf{Frequent Subgraph Mining}. To early prune infrequent patterns, we use the \textit{minimum image-based (MNI) support} \cite{bringmann:frequent} as the frequency of each pattern, which counting the minimum number of distinct mappings for any vertex in the pattern. The support measure is anti-monotonic.
We implement an edge-induced version of FSM in Kaleido.
Given an input graph, edges number of the query pattern $k$ and a threshold of the support, it returns frequent $k$-patterns whose support is beyond the threshold.
To prune infrequent patterns, we loop \texttt{Mapper} and \texttt{Reducer} in each iteration.
Initially, \texttt{Init} calculates the MNI support for each edge ($1$-embedding) and eliminates infrequent edges according to the threshold.
In each iteration of embedding exploration, \texttt{EmbeddingExplorer} expands each embedding by adding a frequent edge. \texttt{EmbeddingFilter(e,<u,v>)} checks if the candidate edge \texttt{<u,v>} is frequent. 
Then \texttt{Mapper} patternizes each embedding, calculates the MNI support for each pattern.
Next, \texttt{Reducer} prunes infrequent patterns and its corresponding embeddings. \texttt{Pattern Filter} eliminates infrequent patterns. 
Finally, in the last iteration, \texttt{Reducer} statistics frequent patterns and returns the results.

\textbf{Motif Counting}. This application counts the frequency of each $k$-motif in the given graph.
The \texttt{EmbeddingFilter} and \texttt{PatternFilter} can be set as default. 
As we know exactly each shape of $k$-motifs like (2 kinds of $3$-motifs, 6 kinds of $4$-motifs and 21 kinds of $5$-motifs, etc.), we stop embeddings generating if $(k-1)$-embeddings are generated. 
Then \texttt{Mapper} explores all canonical $k$-embeddings from each $(k-1)$-embedding, then calculates the hash value of each $k$-embedding.
Finally \texttt{Reducer} aggregates $k$-motifs.
For example, consider counting $3$-motifs over graph in Figure \ref{fig:embgen}.
After the embedding exploration phase, \texttt{EmbeddingExplorer} returns $2$-embeddings of the graph.
In the pattern aggregation phase, considering embedding $s_6$ without loss of generality.
The \texttt{Mapper} explores two $3$-embeddings, $s_{13}$ and $s_{14}$, for $s_6$.
The \texttt{Mapper} obtains a $3$-chain and a triangle.
Finally \texttt{Reducer} aggregates results: 5 $3$-chains and 3 triangles.

\begin{figure}[h]
       \centering
       \includegraphics[width=0.5\textwidth]{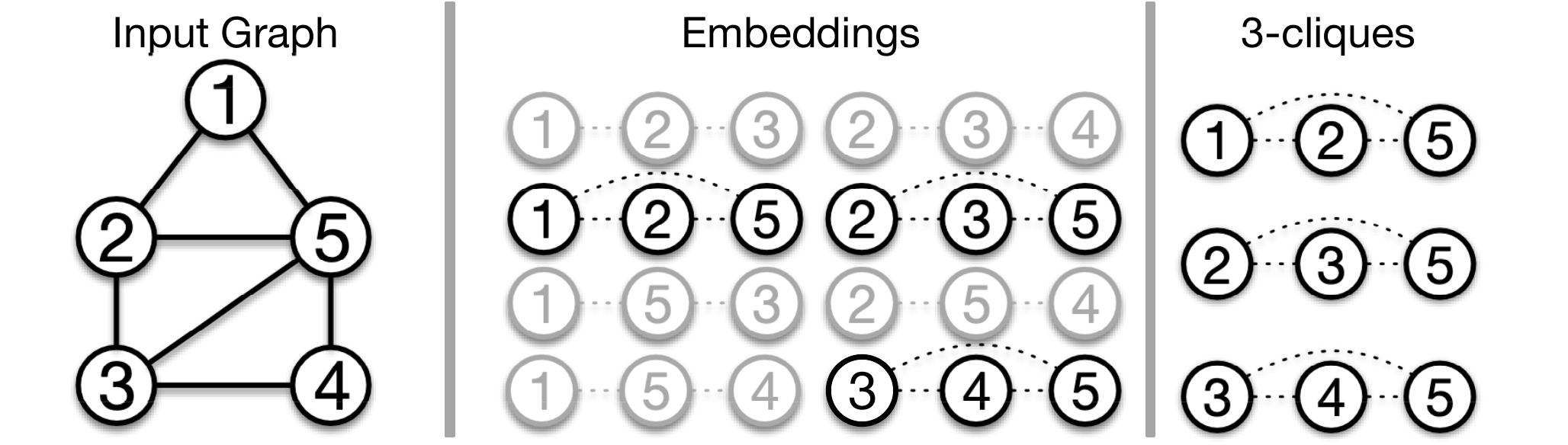}
       \caption{Example of $3$-Clique Discovery. The dotted lines between vertices do not really exist in the embedding array. Gray embeddings indicates that they are eliminated by clique checking filter.}
       \label{fig:3cliq}
\end{figure}

\textbf{Clique Discovery}. This application discoveries all $k$-cliques in the input graph.
\texttt{EmbeddingFilter(e,v)} checks if the candidate vertex $v$ is neighbor of each vertex in embedding $e$.
Therefore generating cases like embedding $\langle 1,2\rangle$ and candidate $3$ will be eliminated, cause vertex $3$ is not neighbor of vertex $1$.
After \texttt{Init}, \texttt{EmbeddingExplorer} prunes illegal embeddings and explores all $k$-cliques after $k-1$ iterations and returns them.
In this example, \texttt{Mapper} is unnecessary to calculate pattern of each embedding, cause all embeddings have a same pattern.
Finally, \texttt{Reducer} return $k$-cliques.

\textbf{Triangle Counting}. This application counts the number of triangles in the input graph.
Initially, \texttt{Init} generates $2$-embeddings from the edge set of the input graph.
Then \texttt{Mapper} counts the number of common neighbors of two vertices in each $2$-embedding canonically.
Finally, \texttt{Reducer} aggregates counting results.
For embedding $s_6=\langle 1,2\rangle$ illustrated in Figure \ref{fig:embgen}, vertex $5$ satisfies former restrictions.
Analogously, cases like vertex $5$ over $s_8$ and $s_{10}$ satisfy the restrictions.
The answer of triangle counting over graph in Figure \ref{fig:embgen} are $3$.

\section{Evaluation}\label{sec:evl}

In this section, we evaluate Kaleido. First we compare Kaleido with the state-of-art graph mining systems, Arabesque and RStream. Then we compare the graph isomorphic checking algorithm in Kaleido with bliss. Next, we test the scalability of Kaleido in different applications. Finally, we test the I/O performance in the hybrid storage. Our experiments are evaluated on a single machine with Intel(R) Xeon(R) Gold 5117 CPU, 128GB memory, and 1 SSD with 480GB disk space. The operating system is CentOS 7.

\subsection{Experimental Setup} \label{subsec:expersetup}
\begin{table}[h]
       \centering
       \caption{Dataset used in evaluation}
       \begin{tabular}{c|c|c|c|c} \hline
       \label{tab:dataset}

       Dataset       & Vertices    & Edges       & Labels      & Avg. Degree \\ \hline \hline
       CiteSeer      & 3,312       & 4,536       & 6           & 3           \\ \hline
       MiCo          & 100,000     & 1,080,298   & 29          & 22          \\ \hline
       Patent        & 3,774,768   & 16,518,948  & 37          & 9           \\ \hline
       Youtube       & 7,065,219   & 59,811,883  & 29          & 17          \\ \hline

       \end{tabular}
\end{table}

\textbf{Datasets}: We use 5 datasets as showed in Table \ref{tab:dataset}. CiteSeer \cite{elseidy:grami} has publications as vertices, with their Computer Science area as a label, and citations as edges. MiCo \cite{elseidy:grami} models the Microsoft co-authorship and consists of an undirected graph whose nodes represent authors and are labeled with the author's field. Patents \cite{leskovec:graphs} includes all citations made by US Patents granted between 1975 and 1999; the year the patent was granted is considered to be the label. Youtube \cite{cheng:statistics} lists crawled video ids and related videos for each video posted from February 2007 to July 2008. The label is the category of each video.

\textbf{Applications}: We test 4 mining applications discussed in Section \ref{sec:api}, FSM, Motif Counting, Clique Discovery and Triangle Counting. For $k$-FSM, we mine the frequent subgraphs which $k-1$ edges and at most $k$ vertices. In our experiments, we run $3$-, $4$-, $5$-FSM over several datasets. Motif Counting executions are run with subgraphs whose number of vertices is 3, 4 or 5. Clique Discovery executions are run with subgraphs whose number of vertices is 3, 4 or 5. Triangle Counting counts the number of triangles in the input graph.

\subsection{Comparisons with Mining Systems}

We compared Kaleido with two state-of-the-art systems, Arabesque \cite{teixeira:arabesque} and RStream \cite{wang:rstream}. We ran all these testing cases in a single node server. 
In testing Arabesque, we deployed the Hadoop 2.7.7 in the experimental environment and put datasets on the local hdfs system, then Arabesque reads input graphs from the hdfs system.
In testing RStream, the partition number of each algorithm was set to 10, 20, 50, 100 respectively, then chose the fastest result. 
We ran algorithms mentioned in Section \ref{subsec:expersetup} over labeled graphs, CiteSeer, MiCo, Patent and Youtube on Kaleido, Arabesque and RStream. 

\begin{table*}[t]
\centering
\caption{Comparisons of running time between Kaleido (KA), Arabesque (AR) and RStream (RS) on four mining algorithms, frequent subgraph mining ($3$-FSM, option: support), motif counting ($k$-Motif, option: $k$), clique discovery ($k$-Clique, option: $k$) and triangle counting (TC) over the former datasets, CiteSeer (CS), MiCo, Patents (PA), Youtube (Ytb). Each result indicate the running time of the application in second. `-' indicates the execution runs out of the memory. `/' indicates the execution runs out of the SSD.}
\label{tab:timeres}
\begin{tabular}{c|c|c|c|c|c|c|c|c|c|c|c|c|c}\hline
\multicolumn{2}{c|}{Dataset}     & \multicolumn{3}{c|}{CS}  & \multicolumn{3}{c|}{MiCo} & \multicolumn{3}{c|}{PA} & \multicolumn{3}{c}{Ytb}\\ \hline
Apps                   & Options & KA                & AR          & RS         & KA                & AR          & RS         & KA                & AR          & RS         & KA                & AR          & RS         \\ \hline \hline
\multirow{4}{*}{$3$-FSM}  & 300  & \textbf{0.04}     & 23.03       & 0.14       & \textbf{7.35}    & 101.77      & 330.72   & \textbf{25.47}    & 139.8       & 1228    & \textbf{132.59}    & 426.67       & -        \\
                          & 500  & \textbf{0.04}     & 17.05       & 0.14       & \textbf{8.19}    & 70.74       & 326.29   & \textbf{26.41}    & 133         & 1220    & \textbf{133.31}    & 409.23       & -        \\
                          & 1000 & \textbf{0.03}     & 17.01       & 0.14       & \textbf{7.84}    & 46.65       & 316.68   & \textbf{28.71}    & 119.4       & 1222    & \textbf{136.24}    & 397.19       & -        \\
                          & 5000 & \textbf{0.02}     & 17.01       & 0.14       & \textbf{3.97}     & 29.67       & 261.70  & \textbf{31.51}    & 102.6       & 1179    & \textbf{155.04}    & 396.11       & -        \\\hline
\multirow{2}{*}{Motif}    & 3    & \textbf{0.03}     & 23.42       & 0.11       & \textbf{1.39}     & 28.37       & 73.92   & \textbf{4.74}     & 79.71       & 100.61  & \textbf{35.47}     & 246.24       & -        \\
                          & 4    & \textbf{0.06}     & 26.10       & 0.42       & \textbf{198.17}   & 284.79      & /       & \textbf{152.28}   & 634.47      & /       & \textbf{4988.96}   & -            & -        \\ \hline
\multirow{3}{*}{Clique}   & 3    & \textbf{0.02}     & 23.01       & 0.02       & \textbf{0.46}     & 27.92       & 4.75    & \textbf{0.56}     & 60.55       & 95.34   & \textbf{2.16}     & 195.63      & -          \\
                          & 4    & \textbf{0.03}     & 27.03       & 0.03       & \textbf{3.88}    & 37.71       & 167.22   & \textbf{1.14}     & 79.68       & 196.37  & \textbf{7.83}     & 461.81       & -         \\
                          & 5    & \textbf{0.04}     & 29.99       & 0.04       & \textbf{183.63} & 299.01       & -        & \textbf{1.46}     & 84.81       & 212.95  & \textbf{18.99}     & 505.9       & -         \\\hline
\multicolumn{2}{c|}{TC}          & \textbf{0.02}     & 23.18       & 0.05       & \textbf{0.17}     & 25.05       & 2.74    & \textbf{0.52}     & 70.17       & 5.4     & \textbf{2.24}     & 287.04       & 39.68     \\\hline

\end{tabular}                
\end{table*}

\begin{table}[h]
\footnotesize
\centering
\caption{Comparisons of memory consumption (MB) between Kaleido, Arabesque (AR) and RStream (RS) over CiteSeer. Each result indicate the memory consumption of the application in Mega-byte.}
\begin{tabular}{c|c|c|c|c} \hline
\label{tab:csmem}
Apps                   & Options & Kaleido                & AR          & RS\\ \hline \hline

\multirow{4}{*}{$3$-FSM}  & 300  & 205.8             & 1916        & \textbf{97.1}\\
                          & 500  & 194.2             & 1888        & \textbf{97.1}\\
                          & 1000 & 130.9             & 1864        & \textbf{97.1}\\
                          & 5000 & \textbf{23.3}     & 1889        & 97.1\\\hline
\multirow{2}{*}{Motif}    & 3    & \textbf{26.9}     & 1802        & 196.7\\
                          & 4    & \textbf{166.8}    & 1812        & 482.5\\ \hline
\multirow{3}{*}{Clique}   & 3    & \textbf{25.8}     & 1890        & 97.1\\
                          & 4    & \textbf{29.7}     & 1932        & 97.1\\
                          & 5    & \textbf{27.2}     & 1940        & 97.1\\\hline
\multicolumn{2}{c|}{TC}          & \textbf{26.9}     & 1819        & 198.3\\
\hline
\end{tabular}                
\end{table}

%

\begin{figure}[h]
       \centering
       \includegraphics[width=0.5\textwidth]{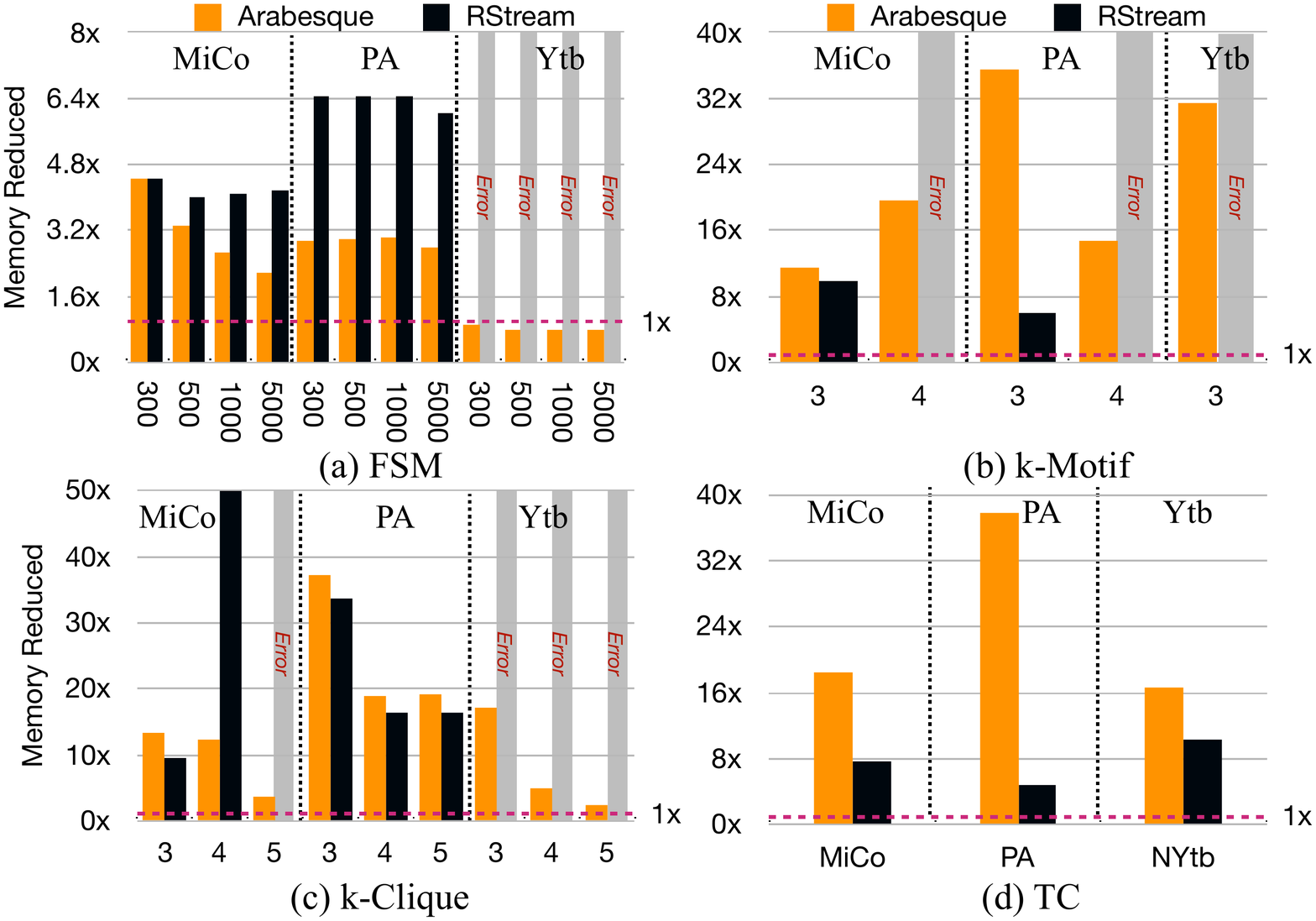}
       \caption{Comparisons of Memory Consumption with Arabesque and RStream. These figures indicate the memory reduction factor of the mining algorithms in Table \ref{tab:timeres}. Each x-axis indicates the argument of each algorithm. The Gray bars indicate the algorithm runs out of the memory.} 
       \label{fig:memtest}
\end{figure}

Table \ref{tab:timeres} reports the running time of the three systems and Figure \ref{fig:memtest} reports the memory consumptions. Note that in this set of experiments, Kaleido and Arabesque run all applications in memory, while RStream writes its intermediate data to disk. Kaleido outperforms both Arabesque and RStream in all cases.
Excluding the small graph CiteSeer, Kaleido outperforms Arabesque by an overall (GeoMean) of \textbf{12.3$\times$} and outperforms RStream by an overall of \textbf{40.0$\times$}; the memory consumption of Kaleido is reduced by \textbf{7.2$\times$} over Arabesque and \textbf{9.9$\times$} over RStream.

Arabesque is a Giraph-based system and implemented by Java. Each iteration of Arabesque is mapped to a superstep of Giraph \cite{avery:giraph}. Arabesque needs extra time to boost the system and fit the basic Giraph API. 
Therefore, for the small graph, CiteSeer, Arabesque is not as efficient as the other two systems. Table \ref{tab:csmem} also indicates that Arabesque allocates a huge amount of memory for the lower layer system. 

RStream is an X-Stream-based system. In the preprocessing phase, RStream partition the input graph into several parts according to an optional partition number given by the user.
RStream uses \texttt{std::set} to maintain the graph topological structure, therefore it fails in loading Youtube in our 128 GB memory environment.
The triangle counting in RStream uses another data structure of the graph and counting strategy, and it runs normally with the GRAS model.

\begin{figure}
    \centering
    \subfloat[Run-Time\label{fig:fsmtrend-a}]{
        \includegraphics[width=0.47\linewidth]{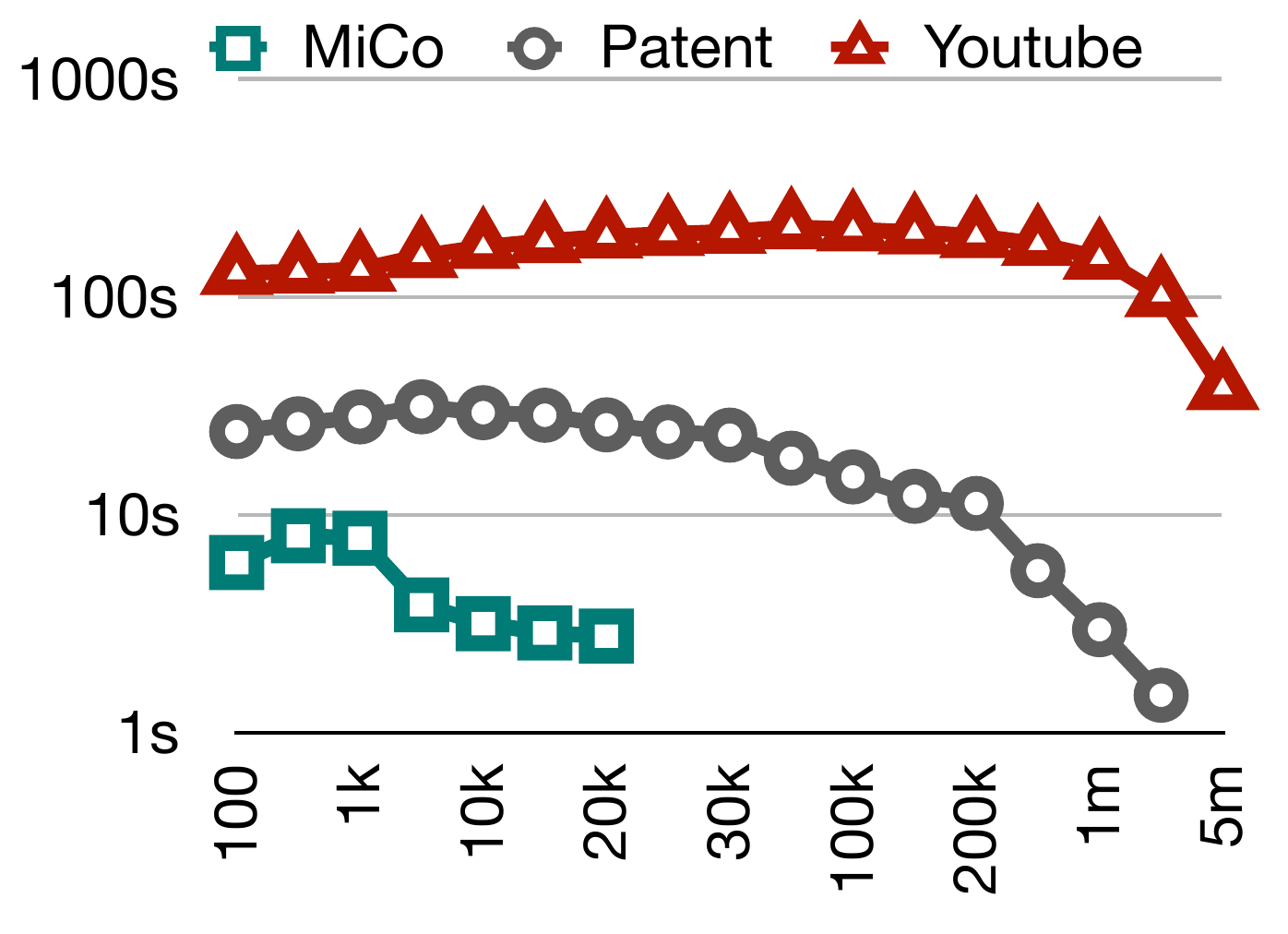}
    }
    \hfill
    \subfloat[Memory Consumption\label{fig:fsmtrend-b}]{
        \includegraphics[width=0.47\linewidth]{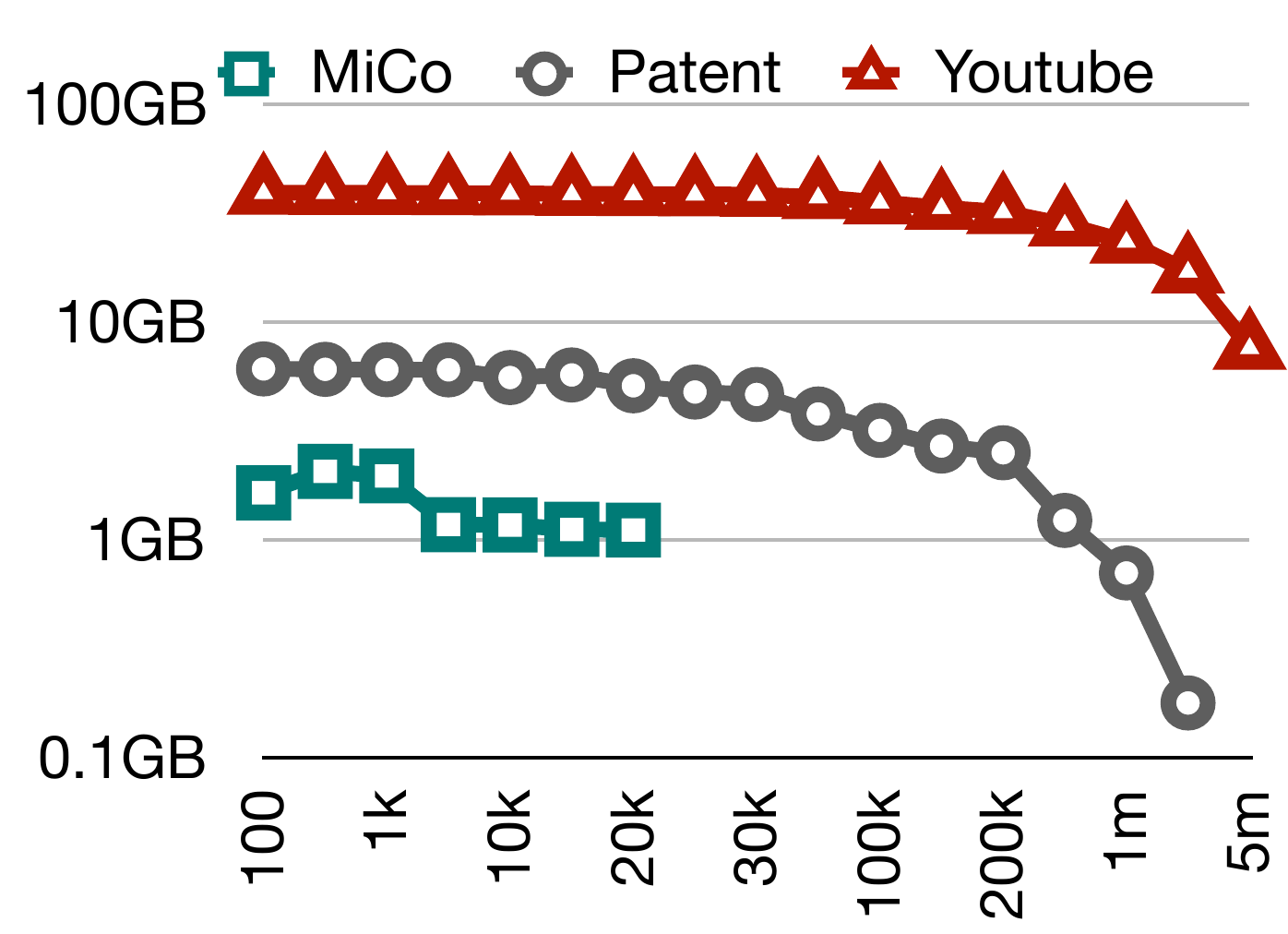}
    }

    \caption{The run-time and the memory consumption trends over the increasing of support in $3$-FSM. The x-axis indicates the support of $3$-FSM.}
    \label{fig:fsmtrend}
\end{figure}

\textbf{FSM}: As discussed earlier, we ran FSM by exploring subgraphs in edge-induced strategy and we used the minimum image-based support metric \cite{bringmann:frequent}, which defines the frequency of a pattern as the minimum number of distinct mappings for any vertex in the pattern, over all instances of the pattern. 
We explicitly state the support used in each experiment, since this parameter is sensitive to the input graph. Theoretically, the smaller support is, the more computation is needed. 
However, the calculation of MNI support for each pattern needs much more computation resources.
In the implementation of the FSM in Kaleido, we do not statistic the accurate MNI support of each pattern. Instead, when the MNI support of any pattern reaches the threshold given by the user, we mark this pattern a frequent pattern and prune it from the candidate.
Therefore the run-time of the FSM computation in Kaleido does not decrease monotonically as the support increases, as illustrated in Figure \ref{fig:fsmtrend}. It will increases to peak time, due to meeting the pruning threshold is getting harder, then decreases normally because the frequent vertices and edges are more and less.


As shown in Table \ref{tab:timeres} and Figure \ref{fig:memtest}a, Kaleido outperforms Arabesque by a GeoMean of 8.5$\times$, 4.4$\times$ and 2.9$\times$ over MiCo, Patent and Youtube respectively, while the memory consumption reduces a GeoMean of 3.1$\times$ and 2.9$\times$ over MiCo and Patent respectively and increases 1.2$\times$ over Youtube.
As discussed in \ref{subsec:embmem}, comparing with the intermediate data structure of Arabesque ODAG, the structure of embeddings CSE in Kaleido saves time from the extra canonical checking when travel the embeddings, but it trades some space of the intermediate data to obtain more efficient performance since the space complexity of ODAG is $O(|V|^2)$.
Even so, Kaleido saves considerable space comparing Arabesque over MiCo and Patent, because Arabesque needs a huge amount of memory to establish its based system and graph data structure and the isomorphism checking library bliss also consumes considerable space.

Comparing with RStream, Kaleido outperforms it by a GeoMen of 46.7$\times$ over MiCo and 43.4$\times$ over Patent, while the memory consumption reduces a GeoMean of 4.2$\times$ over MiCo and 6.3$\times$ over Patent. We found that in the relational phase of RStream \cite{wang:rstream}, the shuffling operation and the aggregating operation produce many memory allocations and deallocations. The shuffling operation turns each tuple into a quick pattern, which allocates and deallocates memory frequently. The aggregating operation builds a hashmap to statistic the support of each pattern in using bliss. We will discuss the comparison with bliss in Section \ref{sec:bliss}.

\textbf{Motif Counting}: Motif counting assumes the input graph is unlabeled and explores all of the embeddings exhaustively until the subgraph reaches the maximum size. 
Table \ref{tab:timeres} and Figure \ref{fig:memtest}b respectively report the comparison of the run-time and the memory consumption with Arabesque and RStream. 
Comparing with Arabesque, Kaleido outperforms by a GeoMean of 6.8$\times$ and the memory consumption reduces by a GeoMean of 20.7$\times$.
Comparing with RStream, Kaleido outperforms by a GeoMean of 33.6$\times$ and the memory consumption reduces by a GeoMean of 7.7$\times$.
$4$-Motif in RStream needs 6 iterations to explore all of $4$-embeddings and writes too much intermediate data to disk, so that our 480 GB SSD cannot afford it.
Thus we tested $4$-Motif in RStream over MiCo and Patent on another server, which has an Intel(R) Xeon(R) E5-2640 v4 CPU with a total of 40 threads, 128 GB RAM and 4 TB Seagate ST4000NM0024-1HT HDD disk.
RStream produces 1.64 TB and 549.15 GB intermediate data over MiCo and Patent respectively and finishes in 114917s and 19740s.

\textbf{Clique Discovery and Triangle Counting}: Clique Discovery is to enumerate all complete subgraphs in the input graph. Triangle Counting is to count the number of triangles in the input graph. 
Table \ref{tab:timeres} and Figure \ref{fig:memtest}c and Figure \ref{fig:memtest}d respectively report the comparison of the run-time and the memory consumption with Arabesque and RStream. 
Comparing with Arabesque, Kaleido outperforms by a GeoMean of 34.1$\times$ and the memory consumption reduces by a GeoMean of 10.6$\times$. Arabesque has a good performance in running $5$-cliques over MiCo. Because MiCo is a denser but smaller graph. The ODAG saves a huge amount of memory while CSE must store many repeating vertices in deeper layers. 
Comparing with RStream, Kaleido outperforms by a GeoMean of 72.0$\times$ and the memory consumption reduces by a GeoMean of 25.0$\times$. To discover $k$-cliques, RStream uses a tricky solution with only $k$ iterations of the edge-induced exploration. However, it still performs than Arabesque except $3$-clique over MiCo and produces many intermediate data. For example, it produces 51.2 GB intermediate data in $4$-clique over MiCo.

\subsection{Comparisons with Isomorphism Checking Algorithms}\label{sec:bliss}

In this section, we compare our isomorphism checking algorithm with the state-of-the-art library, bliss \cite{junttila:bliss}.
To test bliss, we replace the isomorphism checking algorithm in Kaleido with bliss.

\begin{figure}[]
       \centering
       \includegraphics[width=0.5\textwidth]{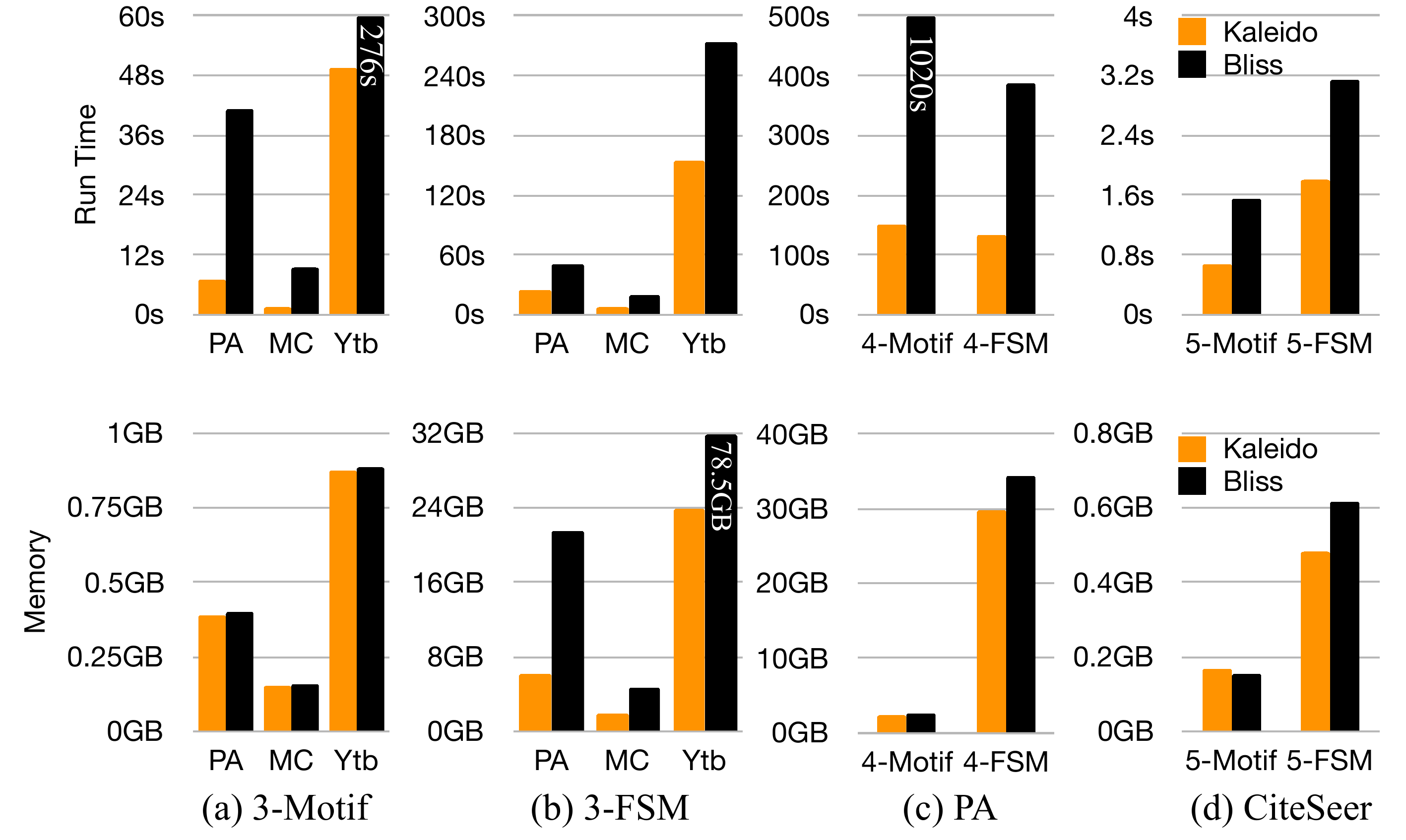}
       \caption{Comparisons of isomorphism checking algorithms with bliss. Figure a and b compare $3$-Motif and $3$-FSM over Patent, MiCo and Youtube. Figure c compares $4$-Motif and $4$-FSM over Patent. Figure d compares $5$-Motif and $5$-FSM over CiteSeer. The upper figures compare the run-time; the lower figures compare the memory consumption} 
       \label{fig:bliss}
\end{figure}

Figure \ref{fig:bliss} illustrates the comparison with bliss.
To fully evaluate Kaleido's isomorphism checking algorithm, we compared $3$-FSM, $4$-, $5$-FSM and Motif Counting respectively over different datasets.
For motif counting, the speedup is 5.8$\times$ but the memory consumption is similar.
For FSM, the speedup is 2.1$\times$ and the memory consumption reduces by 3.1$\times$.
The reason is that the pattern considered by motif counting only contains the structural information of subgraphs, while it contains the label information in FSM.
Kaleido builds the weighted adjacency matrix for each pattern, while Bliss builds search trees.
In FSM, Bliss needs larger hash space and consumes more memory than motif counting.
On the other hand, the counting of motifs is a simple statistic of the occurrence of each motif, while FSM calculates the MNI support of each pattern and this calculation needs quite an amount of computation resources.
Therefore the speedup of replacing Bliss in FSM is not as high as motif counting.

\begin{figure}[h]
       \centering
       \includegraphics[width=0.48\textwidth]{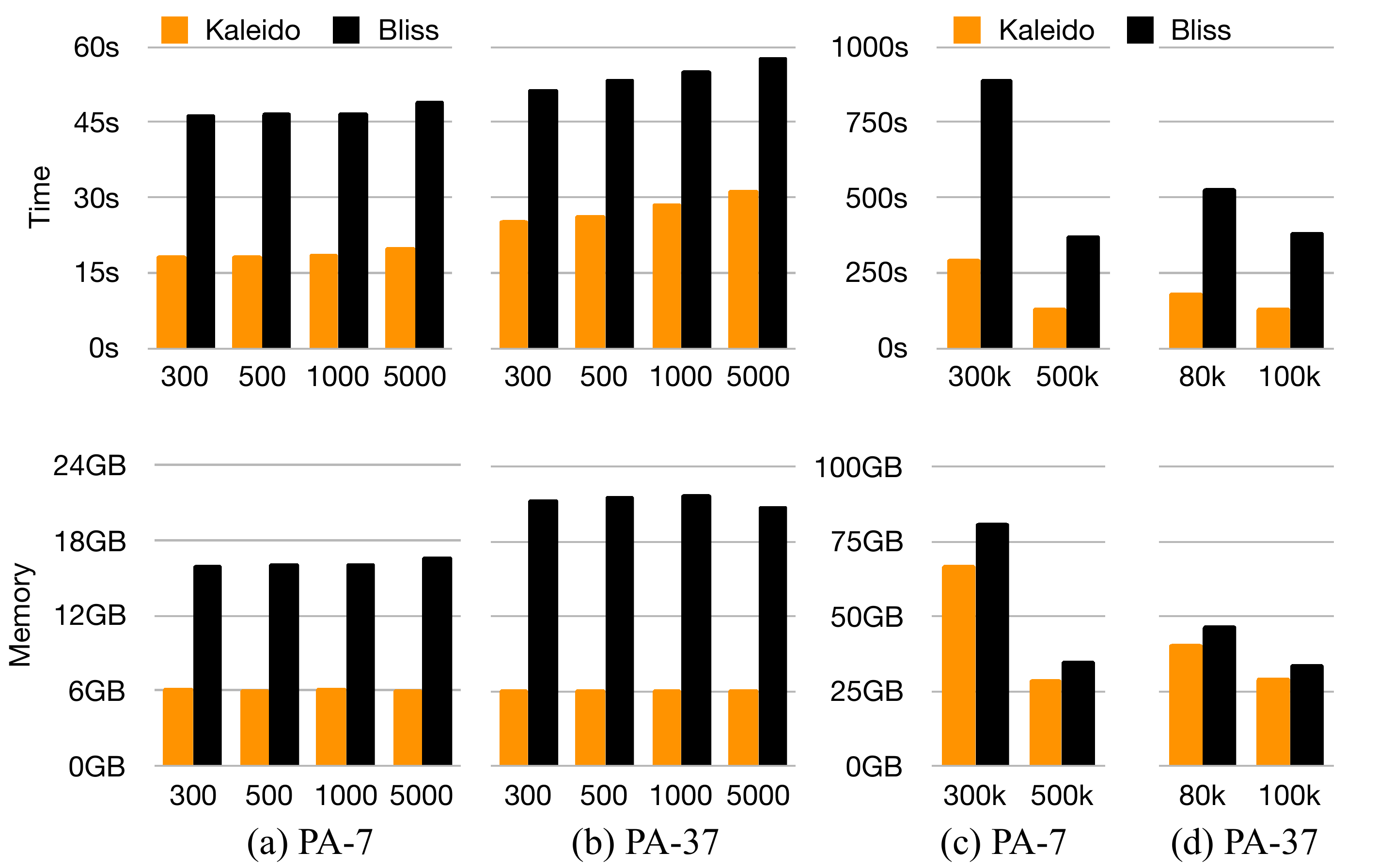}
       \caption{Comparisons of Kaleido and bliss in running $3$-FSM and $4$-FSM when the number of vertex labels changes in 7-label Patent (PA-7) and 37-label Patent (PA-37).
       Figure a and b show results of $3$-FSM; Figure c and d show results of $4$-FSM.
       The x-axis indicates the support of FSM applications.} 
       \label{fig:blisslabel}
\end{figure}

Since the graph Patent possesses two levels of vertex labels, the category and the sub-category of each patent, we tested the performance of $3$-FSM and $4$-FSM over Patent in Kaleido and bliss with several supports, when the number of vertex labels changes.
Figure \ref{fig:blisslabel} illustrates the result.
When the number of vertex labels increases, both Kaleido and Bliss needs more time.
The reason is the implementation of FSM in Kaleido, in which Kaleido does not statistic the accurate MNI support of each pattern but prunes patterns which reach the support threshold from the candidate instead.
When the support threshold is fixed, the 37-label Patent (PA-37) ought to return less frequent patterns than the 7-label Patent (PA-7).
As Figure \ref{fig:fsmtrend} illustrates, the run-time trend of the support in $3$-FSM is to increase first and then decrease.
Note that Figure \ref{fig:fsmtrend} illustrates that the run-time trend is to decrease when the support is larger than 10,000.
Thus Figure \ref{fig:blisslabel}b shows more run-time in PA-37 than Figure \ref{fig:blisslabel}a in PA-7.
The number of frequent embeddings in PA-7 and PA-37 is close (335,781,273 and 335,035,665 respectively in support 300).
It almost needs no extra space in Kaleido, while bliss needs extra memory to build more complexity search trees and maps to a larger hash space.
In testing $4$-FSM, the intermediate embeddings account for the major memory consumption, while Bliss needs longer run-time to check the isomorphism over embeddings.
The number of frequent embeddings $4$-FSM with support 500 K over PA-7 and $4$-FSM with support 100 K over PA-37 is close (1,303,911,410 and 1,490,970,608).
The run-time and the memory consumption of these two applications is close too.
The reason is the support constrains that frequent patterns contain only one kind of vertex label.
It concludes that Bliss is more sensitive with the vertex label information of the input graph than Kaleido; when the number of vertex labels increases, Bliss needs more space to calculate the hash value of each pattern.


\subsection{Scalability}



\begin{figure}[h]
    \centering
    \subfloat[Run-Time\label{fig:scale-a}]{
        \includegraphics[width=0.47\linewidth]{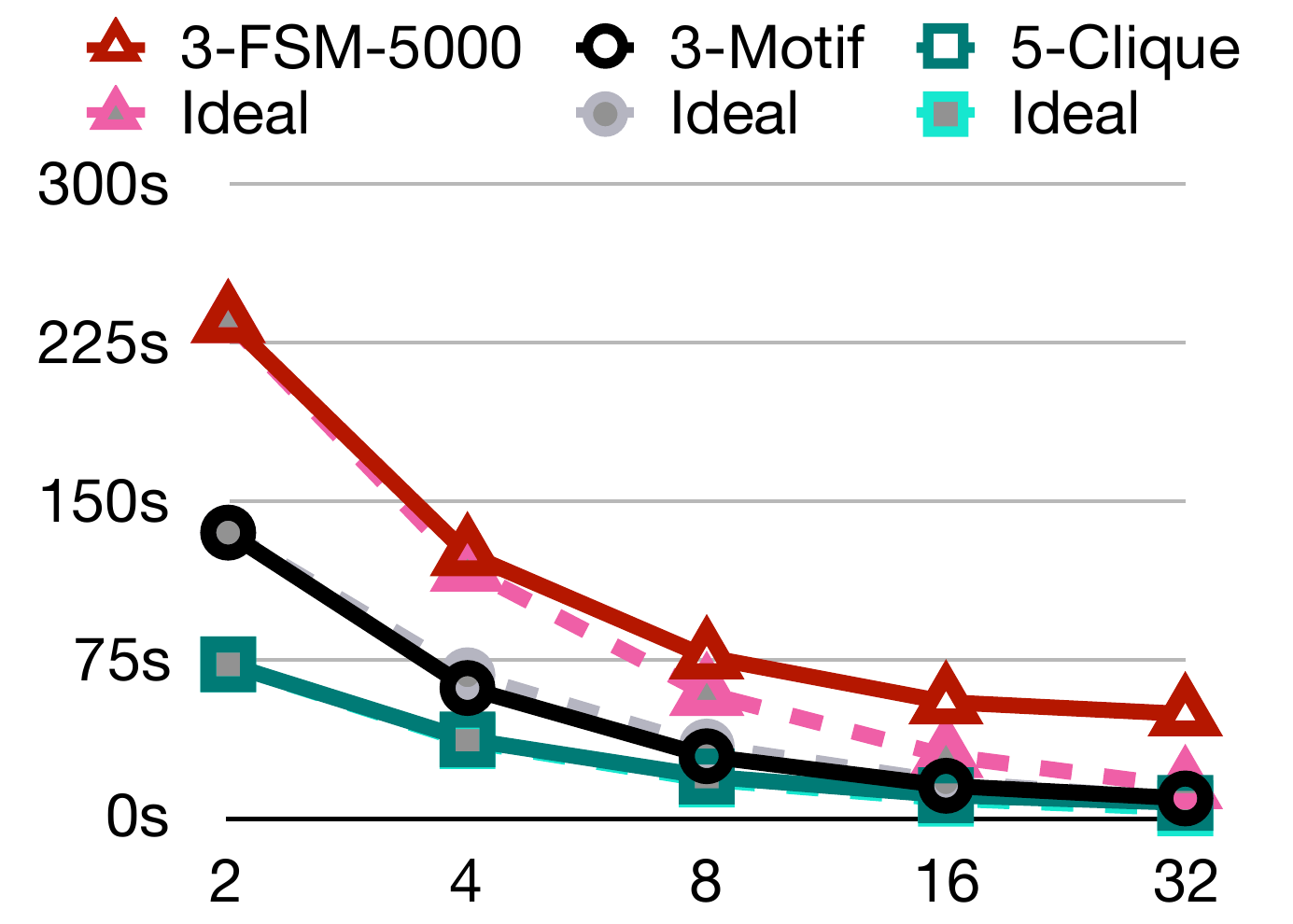}
    }
    \hfill
    \subfloat[Memory Consumption\label{fig:scale-b}]{
        \includegraphics[width=0.47\linewidth]{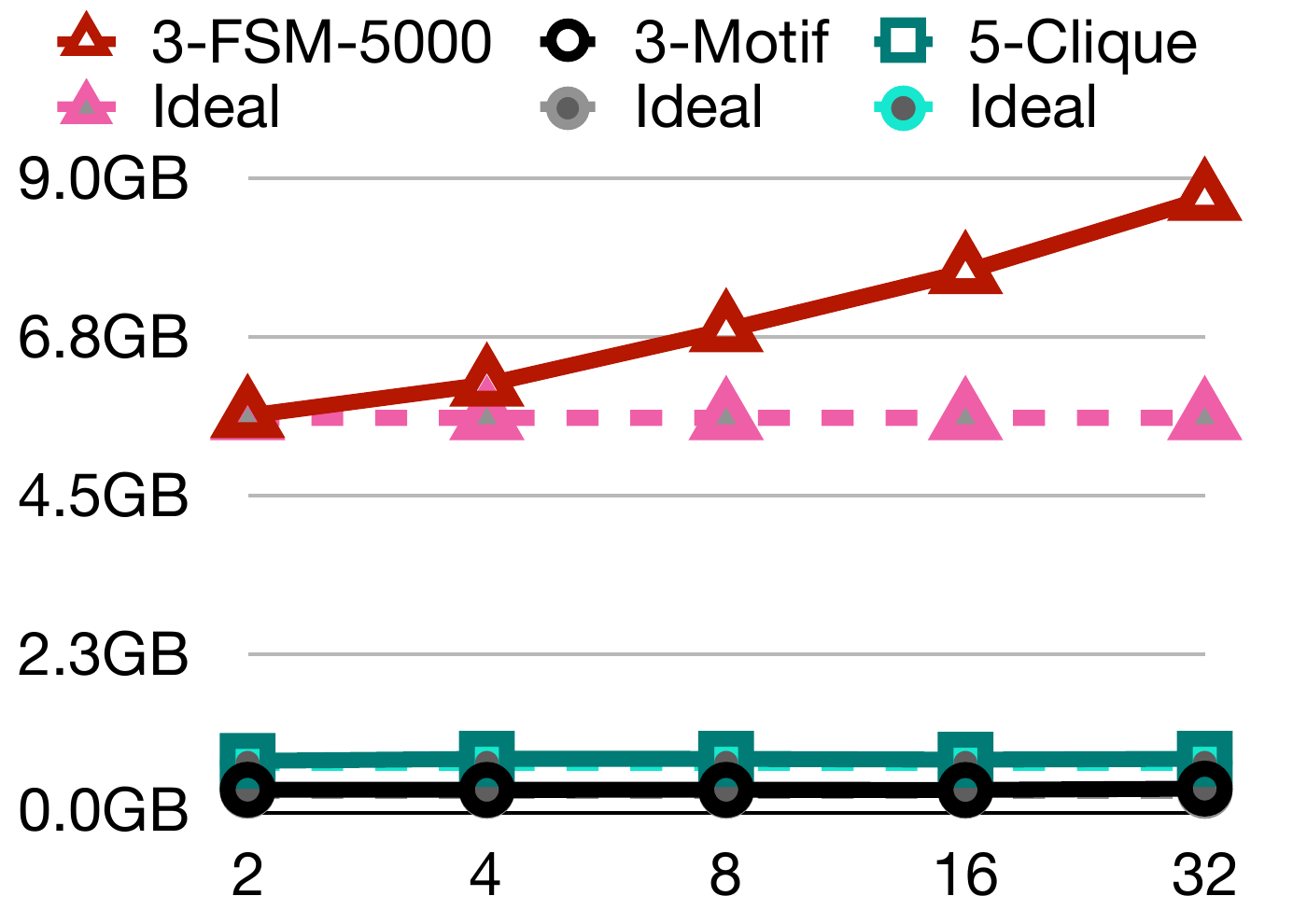}
    }

    \caption{Scalability of Kaleido in 2, 4, 8, 16, 32 threads. Figure \ref{fig:scale-a} shows the run-time; Figure \ref{fig:scale-b} shows the memory consumption. The dotted lines indicate the ideal run-time and memory consumption respectively.}
    \label{fig:scale}
\end{figure}

We tested $3$-FSM with 5000 support, $3$-Motif and $5$-Clique over Patent in varying numbers of threads.
Figure \ref{fig:scale} illustrates Kaleido's run-time and memory consumption for this experiment.
It illustrates that Motif Counting and Clique Discovery scale ideally both in the run-time and the memory consumption.
While FSM only performs sublinearly in the run-time and the memory consumption increases as the number of threads grows.
The implementation of FSM causes this phenomenon. To avoid using concurrent hashmap in the statistic of frequent patterns, we calculate the support of each pattern in every thread independently. 
It avoids the synchronization over each thread, but it consumes more memory in the pattern computation phase.
The overhead of merging aggregating hashmap for FSM in multi-thread is inevitable in our implementation.
If we could replace it by a efficient concurrent hashmap, the scalability of FSM in Kaleido would near linear scaling.

\subsection{I/O and Load-balance in Hybrid Storage}


To evaluate the performance of hybrid storage of the intermediate data, we ran $4$-FSM over Patent with 50k and 100k supports and $4$-Motif over Patent and MiCo in memory and on the hybrid storage respectively. In the hybrid storage testing, we stored the last layer of CSE on SSDs.
Table \ref{tab:outapp} reports the result of these applications. 
It illustrates that the performance attenuation of using hybrid storage in Kaleido is acceptable (lower than 30\% in these applications).
For $4$-FSM over Patent, the memory consumption reduced by the size of the last layer in CSE.
For $4$-Motif, the memory consumption increases, because we built a buffer in fixed size for each thread (in these applications, 16 MB) and the total size of buffers is larger than the last layer of embeddings.
Note that $k$-Motif only stores $k-1$ layers embeddings in Kaleido.

\begin{table}[h]
\centering
\caption{Performance of Kaleido on the hybrid storage in $4$-FSM over Patent with 50k and 100k supports and $4$-Motif over Patent and MiCo.}
\begin{tabular}{c|c|c|c} \hline
\label{tab:outapp}
 Apps                   & In-Memory & Time(s) & Memory (GB)\\ \hline \hline
 \multirow{2}{*}{$4$-FSM(PA,50k)}  & Yes  & 312.1 & 76.7        \\ \cline{2-4}
                           & No  & 362.7 & 15.8 \\\hline
 \multirow{2}{*}{$4$-FSM(PA,100k)} & Yes  & 125.7 & 32.8 \\\cline{2-4}
                           & No  & 135.8 & 11.4 \\\hline
 \multirow{2}{*}{$4$-Motif(PA)} & Yes  & 152.2 & 2.5 \\\cline{2-4}
                           & No  & 249.2 & 2.7 \\\hline
 \multirow{2}{*}{$4$-Motif(MC)} & Yes  & 198.1 & 0.6 \\\cline{2-4}
                           & No & 247.5 & 1.4 \\
                            \hline
\end{tabular}                
\end{table}

%

For $4$-FSM over Patent with 100k support, the memory consumption is 11.4 GB and the size of the intermediate data is less than our experimental server (128 GB).
To fully evaluate the design of embedding hybrid storage in Kaleido, we used cgroup \footnote{A cgroup is a collection of processes that are bound to a set of
limits or parameters defined via the cgroup filesystem. http://man7.org/linux/man-pages/man7/cgroups.7.html} in Linux to limit the maximum RAM of Kaleido in our experimental environment.

\begin{figure}[]
       \centering
       \includegraphics[width=0.5\textwidth]{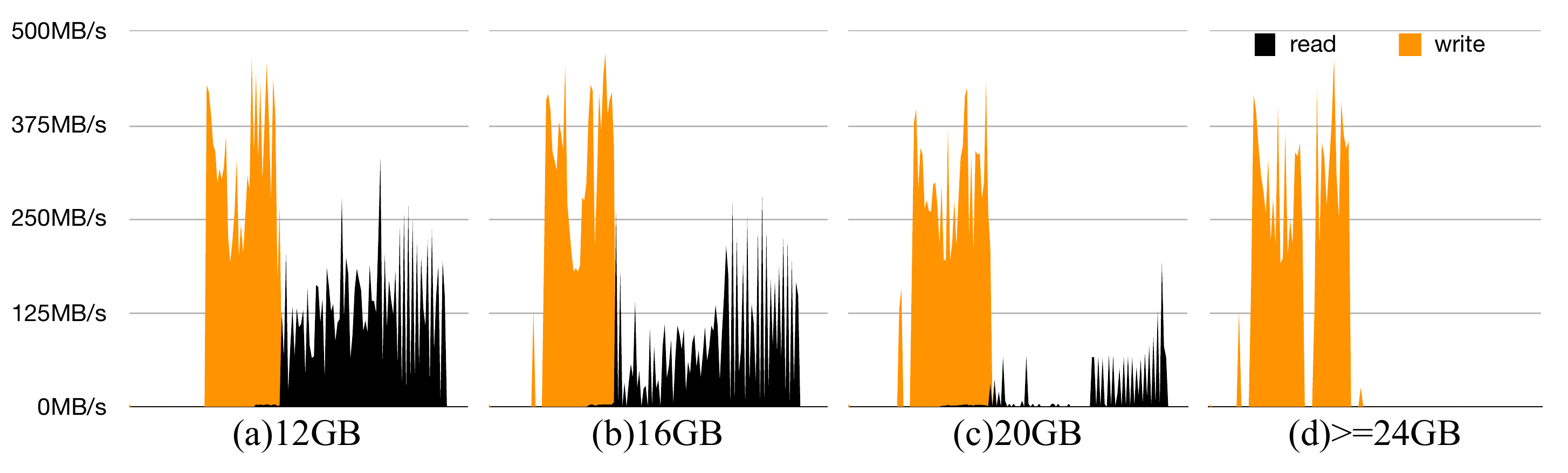}
       \caption{I/O of $4$-FSM over Patent with support 100k. These four figures show the I/O in limiting the memory cache of Kaleido with cgroup. The x-axis indicates the run-time of FSM; the y-axis indicates the reading and writing speed.} 
       \label{fig:iofsm}
\end{figure}


\begin{figure}[]
       \centering
       \includegraphics[width=0.4\textwidth]{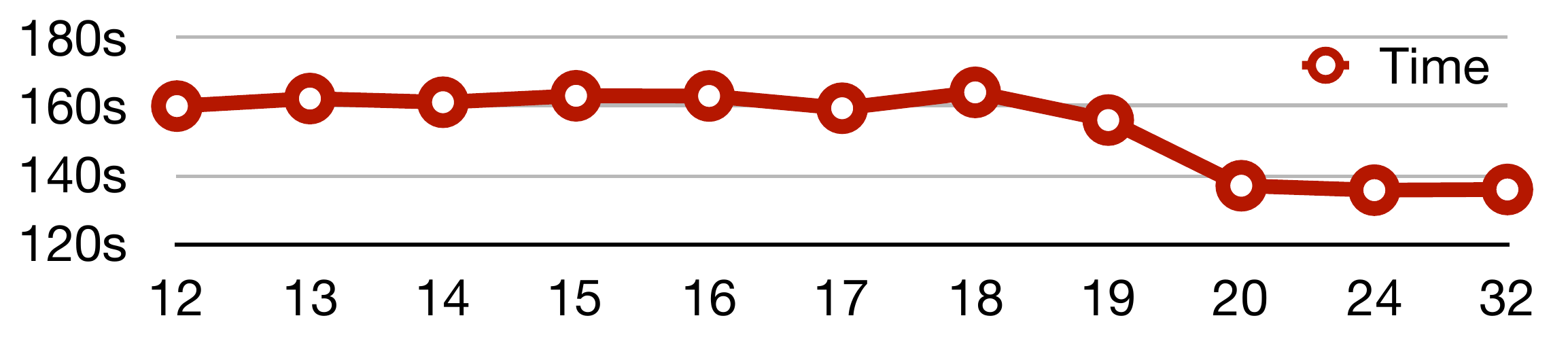}
       \caption{Run-time of $4$-FSM over Patent with 100k support in the different limitation of maximum RAM. The x-axis indicates the limitation of max RAM; the y-axis indicates the run-time.} 
       \label{fig:limit}
\end{figure}

Figure \ref{fig:iofsm} illustrates the I/O of this application in different limitations of max RAM. When the limitation of maximum RAM is larger than 24 GB, the intermediate data will be fully cached in memory. 
Figure \ref{fig:limit} illustrates the run-time of different limitations of max RAM. When the limitation of maximum RAM is lower than 20 GB, the application reads the intermediate data from the disk and the run-time increases within 20\%.

\subsubsection{Load-balance}

\begin{figure}[]
       \centering
       \includegraphics[width=0.45\textwidth]{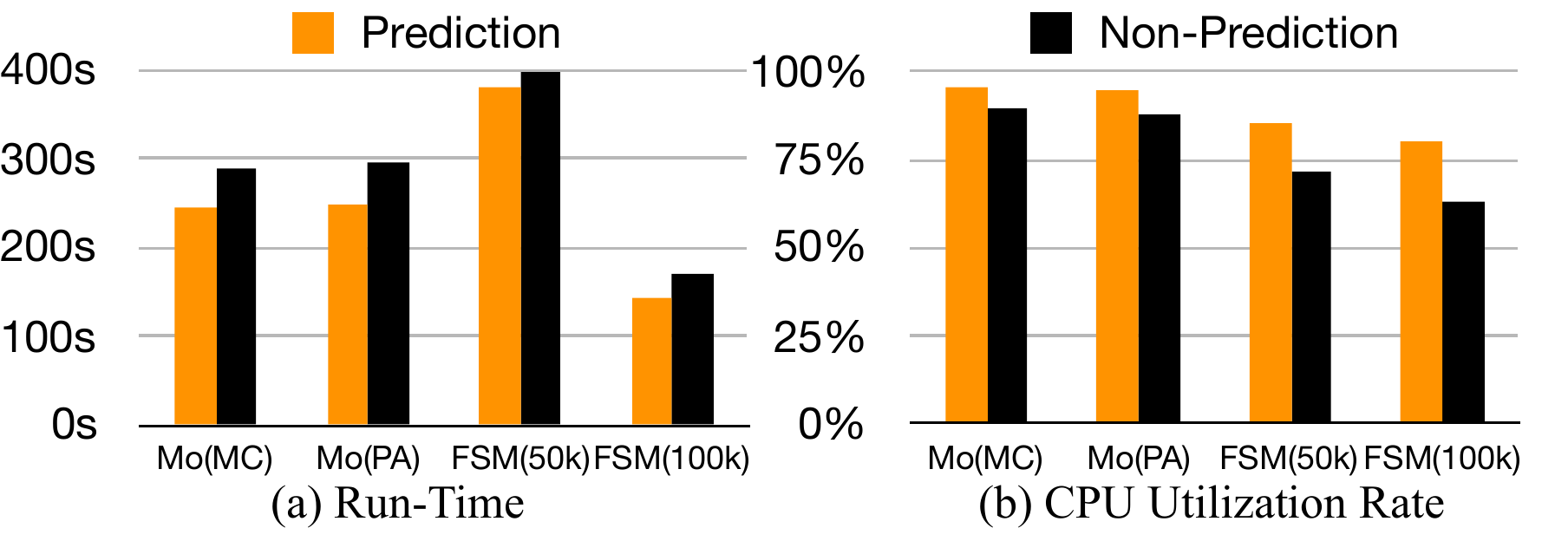}
       \caption{The comparison of prediction and non-prediction in hybrid storage. The first two columns in Figure a and b compare the $4$-Motif over MiCo and Patent; the last two columns compare the $4$-FSM over Patent with supports 50 K and 100 K.} 
       \label{fig:balance}
\end{figure}

\begin{figure}[]
       \centering
       \includegraphics[width=0.45\textwidth]{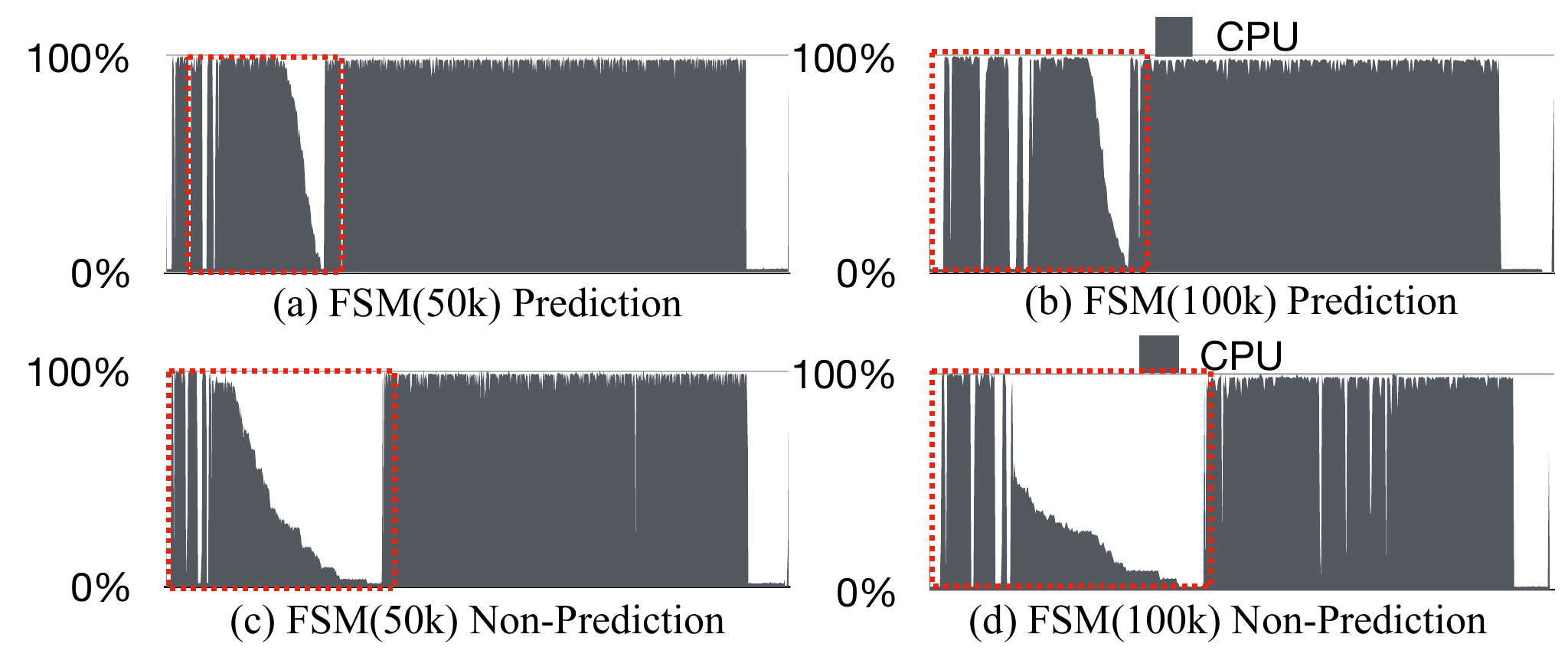}
       \caption{Comparisons of the CPU utilizing rate in $4$-FSM over Patent with supports 50 K and 100 K. The dotted boxes indicate the embedding exploration phase in FSM.} 
       \label{fig:cpu-ba}
\end{figure}

We evaluated the load-balance in hybrid storage by verifying the effectiveness of the prediction of the candidate size.
We ran $4$-FSM with support 50 K and 100 K over Patent and $4$-Motif over Patent and MiCo.
The result is illustrated in Figure \ref{fig:balance}.
The prediction of the candidate size saves the run-time in the exploration over hybrid storage and it outperforms 1.2$\times$ over non-prediction.
Figure \ref{fig:cpu-ba} illustrates the CPU utilizing rate of $4$-FSM over Patent with supports 50 K and 100 K and the prediction improve the efficiency of the exploration significantly.

\section{Related Work}\label{sec:rw}

Over the last decades, graph mining has emerged as an important research topic. Here we discuss the state-of-the-art for the graph mining problems tackled in this paper.

\textbf{Graph Mining Algorithms}
gSpan \cite{yan:gspan} is an efficient frequent subgraph mining algorithm designed for mining multiple input graphs.
However, gSpan is designed for multiple graphs of mining problems. If we have a single input graph, we have to find multiple instances in the same graph, therefore it complexes the problem.
Michihiro \emph{et al.} \cite{kuramochi2004finding} first proposed algorithms to mine patterns from a single graph.
They use an expensive anti-monotonic definition of support based on the maximal independent set to find edge-disjoint embeddings.
GraMi \cite{elseidy:grami} proposes an effective method in the single large graph and presents an extended version with supporting structural constraints and an approximate version.
Pr{\v{z}}ulj \emph{et al.} \cite{prvzulj:biological} introduces the motif counting problem.
Ribeiro \emph{et al.} \cite{ribeiro:gtries} presents G-Tries which is an effective approach for storing and finding the frequency of motifs.
Apar{\'\i}cio \emph{et al.} \cite{aparicio:parallel-gtries} designs and implements a parallel version of G-Tries.
Maximal clique is well studied problem.

\textbf{Graph Mining Systems}
Arabesque \cite{teixeira:arabesque} is a distributed graph mining system which supports popular mining algorithms. Arabesque proposed a graph exploration model with the concept of embeddings. Arabesque explores all the embeddings under constraining of user-defined filters and the developer processes each embedding with a filter-process programming model. Compared with Kaleido, Arabesque needs another canonically checking of each embedding in traveling embeddings. 
ScaleMine \cite{abdelhamid:scalemine} is a parallel frequent subgraph mining system, which computes the approximate solution of frequent patterns firstly and statistics the exact solution by using the results of the first step to prune the search space.
NScale \cite{quamar:nscale} is designed to solve graph mining problems using MapReduce framework. It proposes a neighborhood-centric model, in which a $k$-hop neighborhood subgraph of an interest-point is constructed with $k$ rounds of Map-Reduce and each round of Map-Reduce extends the $1$-hop new neighbors. However, the overhead of MapReduce in processing candidate subgraphs is very high.
G-Miner \cite{chen:gminer} is a distributed graph mining system, which models subgraph processing as independent tasks and designs suitable scheduling for the task pipeline. However, G-Miner does not support FSM and motif counting.
ASAP \cite{iyer:asap} is a distributed, sampling-based approximate computation engine for graph pattern mining. ASAP leverages graph approximation theory and extends it to general patterns in a distributed setting. It allows users to trade-off accuracy for result latency. However, ASAP only counts the interest of the user with an acceptable error, like motif counting and pattern matching, but cannot return the exact result of frequent patterns.
RStream \cite{wang:rstream} is the first single-machine, out-of-core graph mining system. RStream employs a GRAS programming model which combines GAS model and relational algebra to support a wide variety of mining algorithms. However, the join of subgraphs and edges of the input graph in RStream is still an expensive operation. The edge-induced exploration of subgraphs also complexes some mining problems, like motif counting and clique discovery.

\textbf{Graph Isomorphism Checking Libraries} 
The most common practical approach for the graph isomorphism problems is canonical labeling, a process in which a graph is relabeled in such a way that isomorphic graphs are identical after relabeling.
The mainly strategy of the canonical labeling is building search tree for the input graph.
Nauty \cite{mckay1978computing:nauty} is the first program that could handle large automorphism groups; it uses automorphism to prune the search in testing automorphism.
Nauty generates the search tree in depth-first order, while Trace \cite{piperno2008search:trace} introduces a breadth-first search in generating the search tree.
Saucy \cite{darga2004exploiting:saucy} is an implementation of the Nauty system by utilizing the sparsity and particular construction of colored graphs.
However, these libraries focus on the checking of the automorphism, which only suits for the unlabeled graphs.
Bliss \cite{junttila:bliss} supports the isomorphism checking of labeled graphs.
However, building the search tree brings frequently memory allocating and deallocating which slow down the processing and consume a huge amount of memory.
In addition, bliss is designed for the large graph isomorphic checking, while the eigenvalue checking strategy is sufficient in the mining scenes.

\section{Conclusion}\label{sec:conclusion}

In this paper, we present Kaleido, a single-machine, out-of-core graph mining system. 
Kaleido follows the subgraph-centric model and provides a user-friendly simple API that allows non-experts to build graph mining workloads easily. 
To efficiently store and process the huge amount of intermediate data, Kaleido builds a succinct intermediate data structure and adjusts the storage in memory or out-of-core smoothly according to the scale of intermediate data. 
Kaleido designs an lightweight and efficient graph isomorphism checking algorithm for small graphs in which the number of vertices is less than 9. 
Experimental results demonstrates that Kaleido is more efficient than the state-of-the-art graph mining systems in most cases.
The isomorphism checking algorithm in Kaleido is more efficient and consumes less memory than the state-of-the-art graph library.

\bibliographystyle{abbrv}
\balance
\bibliography{GM}  



       

\end{document}